\documentclass[journal,10pt,a4paper,twocolumn,twoside]{IEEEtran}

\IEEEoverridecommandlockouts
\usepackage{adjustbox}
\usepackage{comment}
\usepackage{cite}
\usepackage{physics}
\usepackage{caption} 
\usepackage{amsmath,amssymb,amsfonts,amsthm}
\usepackage{mathtools}
\usepackage{algorithmic}
\usepackage{graphicx}
\usepackage{textcomp}
\usepackage{tikz}
\usetikzlibrary{automata, positioning, arrows,calc}
\usetikzlibrary{quantikz}
\usepackage{algorithm}
\usepackage{algorithmic}
\usepackage{float}
\usepackage{subcaption}
\usepackage[T1]{fontenc}
\usepackage{hyperref}
\usepackage{multirow}
\usepackage{booktabs}
\usetikzlibrary{decorations.pathreplacing}
\definecolor{myLightGray}{RGB}{191,191,191}
\definecolor{myGray}{RGB}{160,160,160}
\definecolor{myDarkGray}{RGB}{144,144,144}
\definecolor{myDarkRed}{RGB}{167,114,115}
\definecolor{myRed}{RGB}{255,58,70}
\definecolor{myGreen}{RGB}{0,255,71}

\newtheorem{cor}{Corollary}

\newtheorem{prop}{Proposition}

\theoremstyle{remark}
\theoremstyle{definition}

\newtheorem*{rem}{Remark}
\newtheorem{defin}{Definition}

\begin{document}

\def\myvdots{\ \vdots\ }

\title{Quantum MAC: Genuine Entanglement Access Control via Many-Body Dicke States}
\author{Jessica Illiano, Marcello~Caleffi$^*$,~\IEEEmembership{Senior~Member,~IEEE}, Michele Viscardi, Angela~Sara~Cacciapuoti,~\IEEEmembership{Senior~Member,~IEEE}
    \thanks{The authors are with the \href{www.quantuminternet.it}{www.QuantumInternet.it} research group, University of Naples Federico II, Naples, 80125 Italy. }
    \thanks{A preliminary version of this work has been presented at IEEE GLOBECOM 2022 \cite{IllVisKou-22}.}
   \thanks{Marcello Caleffi acknowledges PNRR MUR project RESTART-PE00000001, Angela Sara Cacciapuoti acknowledges PNRR MUR project NQSTI-PE00000023.}
}

\newcommand{\eqdef}{\stackrel{\triangle}{=}}

\maketitle
\begin{abstract}
Multipartite entanglement plays a crucial role for the design of the Quantum Internet, due to its peculiarities with no classical counterpart. Yet, for entanglement-based quantum networks, a key open issue is constituted by the lack of an effective \textit{entanglement access control} (EAC) strategy for properly handling and coordinating the quantum nodes in accessing the entangled resource. In this paper, we design a \textit{quantum-genuine} entanglement access control (EAC) to solve the contention problem arising in accessing a multipartite entangled resource. The proposed quantum-genuine EAC is able to: i) fairly select a subset of nodes granted with the access to the contended resource; ii) preserve the privacy and anonymity of the identities of the selected nodes; iii) avoid to delegate the signaling arising with entanglement access control to the classical network. We also conduct a theoretical analysis of noise effects on the proposed EAC. This theoretical analysis is able to catch the complex noise effects on the EAC through meaningful parameters.
\end{abstract}


\section{Introduction}
\label{sec:1}

A fundamental role in the Quantum Internet \cite{CacCalTaf-20,IllCalMan-22,KozWehVan-22,WehElkHan-18} is played by multipartite entanglement \cite{RiePol-11,NieChu-11}, since it enables computing and communication functionalities with no counterpart in the classical world \cite{RauBri-01,Nie-04,RamPirDur-21,VanSatBen-21,VanTouHor-11,CaccIllKou-22,IllCalMan-22}, including (but not limited to) advanced forms of privacy and anonymity \cite{KhaKhaReh-22}, and the so-called \textit{on-demand connectivity} \cite{IllCalMan-22}. 

\begin{figure*}[t]
    \centering
    \begin{subfigure}{0.48\textwidth}
        \centering
         \resizebox{0.7\textwidth}{!}{
        \input{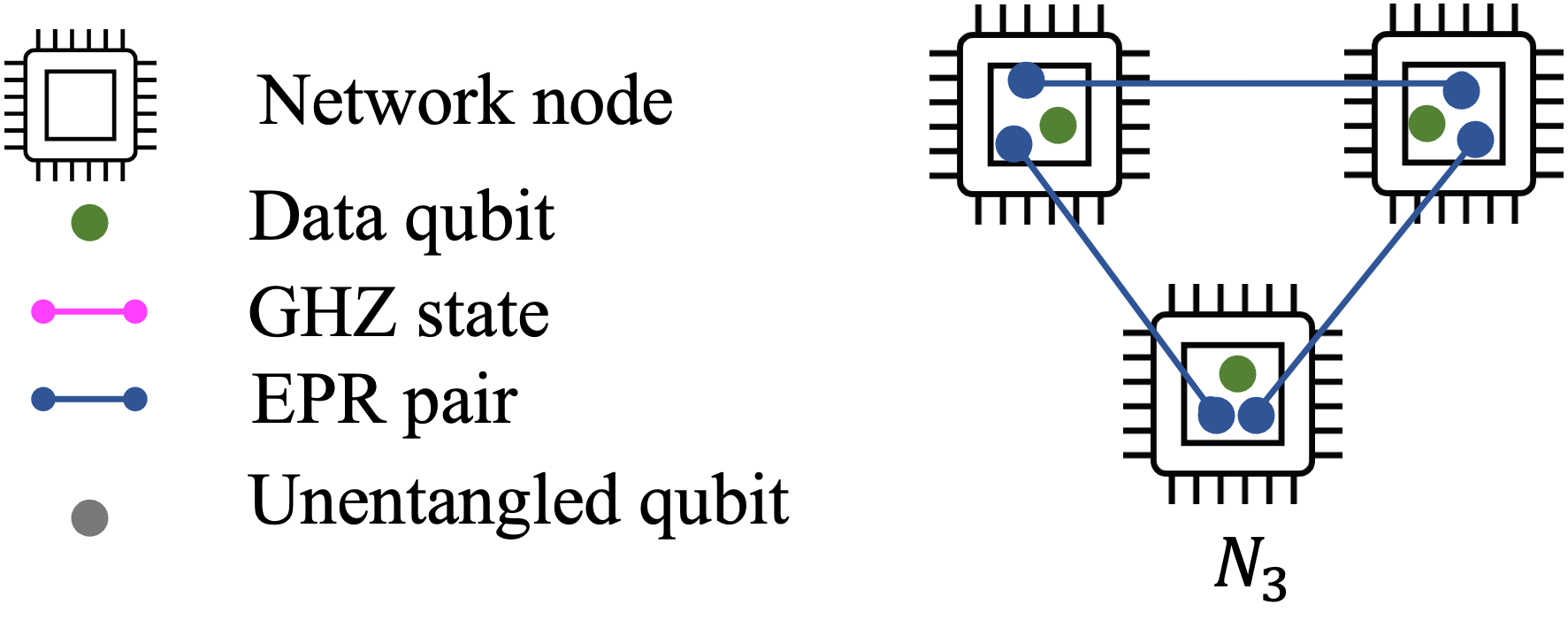}
            }
        \subcaption{\footnotesize{EPR-based connectivity. A minimum of three EPR pairs must be generated and distributed. Once distributed, each EPR establishes a virtual quantum link between a fixed pair of nodes.}}
        \label{fig:01.a}
    \end{subfigure}
    \hfill
    \begin{subfigure}{0.48\textwidth}
    	\centering
        \resizebox{\textwidth}{!}{
        \input{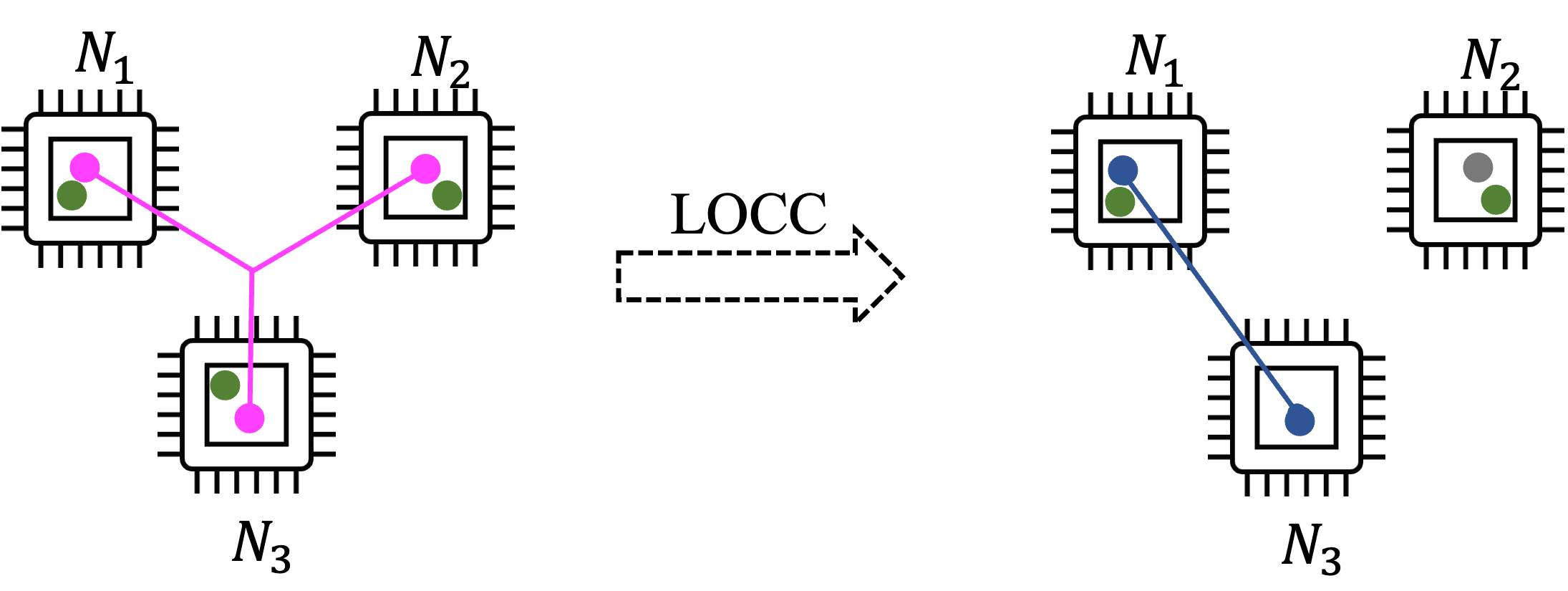}
            }
        \centering
    	\subcaption{\footnotesize{On-demand connectivity. A direct connection between any pair of nodes can be obtained by distributing a multipartite state through local operation and classical communications (LOCC).}}
    	\label{fig:01.b}
    \end{subfigure}
    \caption{\textit{A-priori} vs \textit{on-demand connectivity}.}
    \label{fig:01}
    \hrulefill
\end{figure*}

\begin{figure*}[t]
    \centering
    \resizebox{!}{4cm}{
        \input{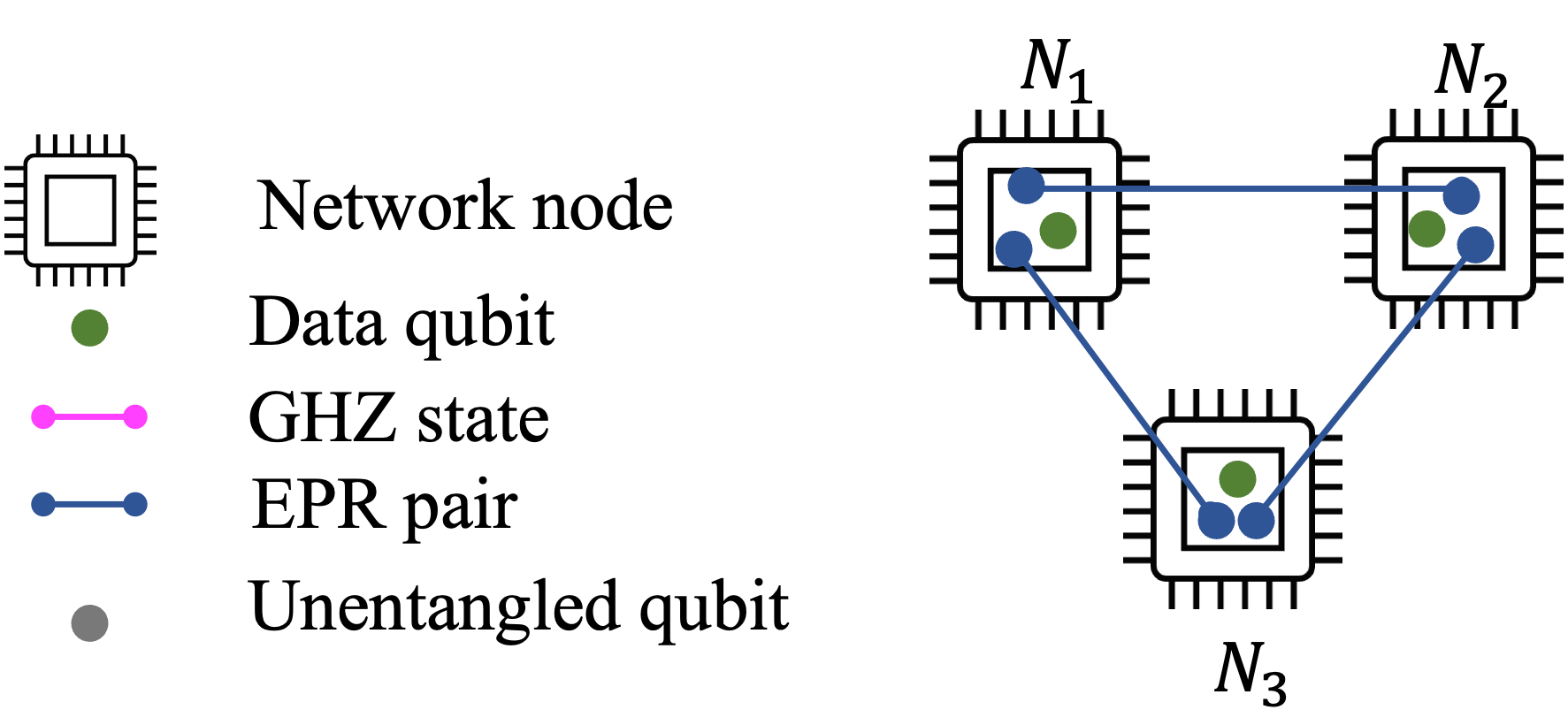}
    }
    \caption{MAC vs EAC. Entanglement Access Control (EAC) is the quantum-equivalent problem of Medium Access Control (MAC). Specifically, while classical-network nodes must coordinate via some MAC protocol to access to a shared channel, quantum-network nodes must coordinate as well via some EAC protocol to access to a shared entangled resource.}
    \label{fig:03b}
    \hrulefill
\end{figure*}

Specifically, as a pivotal example for the on-demand connectivity, let us consider three nodes, say nodes $N_1$, $N_2$ and $N_3$ as in Figure~\ref{fig:01}. In EPR-based networks \cite{VanTou-13,Weh-19,PomDonWeh-21}, to establish a direct\footnote{The term ``direct connectivity'' refers to the availability of an EPR pair shared between two nodes -- regardless of how this EPR has been distributed to the nodes, i.e., through either direct quantum link or multi-hop quantum path -- without the need of any additional \textit{helper} such as an intermediate third node implementing entanglement swapping. Accordingly, the EPR pair can be straightly exploited -- as instance, for quantum teleportation -- by the two nodes without involving any additional node. As detailed in \cite{IllCalMan-22}, such a direct connectivity is related to entangled states shared among remote nodes. However, physical direct connectivity is envisioned as well in quantum networks \cite{LonPanYu-22}.} connectivity among any pair of the node set with cardinality $n$, $\frac{n}{2} (n -1)$ EPR pairs must be properly shared by the nodes, reserving $n-1$ specialized qubits, aka communication qubits \cite{KozWehVan-22,CalAmoFer-22}, in each node to this purpose. Specifically, with reference to Figure~\ref{fig:01}, the EPRs must be distributed so that each node -- say $N_1$ -- reserves two communication qubits, one for each different EPR pairs shared with a different node -- i.e, $N_2$ and $N_3$, respectively. Accordingly, the identity of the nodes that can exploit entanglement as a \textit{communication resource} is fixed a-priori, with no possibility of adapting to time-varying communication needs.
Differently, by considering multipartite-entanglement networks \cite{RamPirDur-21}, multipartite entangled states --  such as GHZ states \cite{GreHorZei-89,IllCalMan-22} -- enable the extraction of an EPR pair between any pair of nodes \textit{at run-time}, depending on the instantaneous communication needs. This key feature enables full connectivity among $n$ nodes, without (unreasonably) requiring a number of communication qubits at each node scaling with $\mathcal{O}(n)$. This constitutes a very attractive feature, since limiting the number of communication qubits decreases the hardware complexity of the nodes and it helps in handling the trade-off between communication and data qubits \cite{KozWehVan-22}. With reference to the example of Figure~\ref{fig:01}, by distributing a 3-qubit GHZ state through the network with one communication qubit at each node, an EPR pair can be extracted at a run time by any pair of nodes, with the identities of the entangled nodes chosen at run-time. From a communication engineering perspective, a key open issue in multipartite-entanglement networks is constituted by the need of proper management and coordination among the entangled nodes, since they all \textit{share} the same multipartite state. Specifically, to leverage the multipartite entanglement advantages, there must be a tight cooperation between the involved network nodes – nodes that must be aware of each other identities – for being able to exploit the quantum correlation provided by entanglement \cite{IllCalMan-22}. Indeed, any processing of a single entangled qubit has an instantaneous effect on the global entangled state, with possible changes affecting the remaining entangled qubits as well, regardless of the distances among the entangled nodes. As pictorially represented in Fig.~\ref{fig:03b}, these network nodes also compete among each others to use the same shared entanglement resource for a targeted application. Hence, when it comes to the design of the network functionalities, proper access to the multipartite entanglement resource is pivotal. As a consequence, an \textit{entanglement access control} (EAC) functionality is  mandatory \cite{IllCalMan-22}. However, so far, EAC has been poorly investigated, by implicitly delegating it to some forms of classical signaling through classical Internet. This, in turn, requires a \textit{functional} classical-quantum interface between the classical Internet and the Quantum Internet, which is still an open issue \cite{IllCalMan-22,CaccIllKou-22}. Thus, limiting classical signaling represents both an attractive strategy accordingly to the current state-of-the-art as well as a yet-to-be solved research problem.

In this paper we address this issue, by designing a ``\textit{quantum-genuine}'' EAC, which abstains from delegating the solution of the entanglement contention to the classical network, by relying on a singled out multipartite entangled state, referred to as ``\textit{contention-resolution state}''. More into details, the proposed EAC exploits the feature of the contention-resolution state to solve the contention problem arising in accessing a multipartite entangled resource, referred to in the following as ``\textit{contended resource}''. By exploiting the features of the designed \textit{contention-resolution state}, the proposed quantum-genuine EAC is able to: i) fairly select a subset of nodes granted with the access to the contended resource; ii) preserve the privacy and anonymity of the identities of the selected nodes; iii) avoid to delegate the signaling arising with entanglement access control to the classical network.  As a matter of fact, the contribution of this paper is not limited to the EAC design. Indeed, we also conduct a theoretical analysis about the quantum noise effects on the performance of the proposed EAC. To this aim, we first recognize that a very crucial noise source affecting the proposed EAC is constituted by the entanglement distribution process, unreliable in providing the network nodes with the required entanglement resources. Stemming from this, we develop a theoretical framework able to capture noisy entanglement distribution processes. Specifically, we prove that a noisy entanglement distribution process can be modeled as a discrete Markov chain, for which we also analytically derive closed-form expressions of the transition probabilities. This is a key result, since it provides a powerful analytical framework that can be exploited beyond the scope of this manuscript for analyzing entanglement-based networks. Then, by accounting for such an analytical framework, we conduct a theoretical analysis of the EAC contention-resolution capabilities in presence of quantum noise. This theoretical analysis is able to catch the complex noise effects on the EAC through meaningful parameters.

\subsection{Related Works}
\label{sec:1.1}

In\cite{BugCouOma-23}, the authors propose a scheme enabling the generation of distributed multipartite GHZ states over remote network nodes through pre-shared EPR pairs. Similarly, in \cite{WalZweMus-16,DebOuyGoo-20,AviRozWeh-22} the authors focus on the aforementioned GHZ state distribution, by considering different network architectures, ranging from centralized to quantum-repeater-based ones. In the same research line discussed above, \cite{PirWarDur-18,MeiMarGro-19,FisTow-21} should also be categorized, but with reference to the wider class of graph states. Additionally, several works focus on protocols for end-to-end EPR distribution, i.e., the distribution of EPR pairs through entanglement swapping and quantum repeaters\cite{Cal-17,PanKroTow-19,ShoQia-20}. With reference to the performance analysis of quantum communication systems, interesting insights are given in \cite{VarGuhNai-21-2,VarGuhNai-21-1}. Specifically, the authors consider a star-network topology with a central node -- referred to as \textit{switch} -- acting as a quantum repeater. By assuming the availability of infinite coherence time and infinite resources at the switch, the authors analyze the expected capacity in terms of number of stored qubits by exploiting statistical tools. In \cite{GyoImr-19}, the authors analyze the entanglement access problem in the light of accessing to a point-to-point EPR pair, with the aim to share end-to-end EPR pairs between two remote nodes with a certain fidelity. Specifically, the authors consider the so-called \textit{entanglement routing problem}, namely, the problem of obtaining end-to-end EPRs between remote nodes by performing entanglement swapping over a multi-hop path composed by point-to-point pre-shared EPRs. To this aim, the authors analyze the probability that two remote nodes are able to exploit (access) a set of point-to-point EPRs.

Differently from all the aforementioned works and to the best of authors’ knowledge, this manuscript is the first work addressing the design of a \textit{quantum-genuine EAC}, able to distributively solve the entanglement contention arising among the network nodes by accounting for the key requirements detailed in Sec.~\ref{sec:3.2}.  

The rest of this manuscript is organized as follows. In Sec.~\ref{sec:2}, we describe the system model and we collect some definitions utilized through the paper. In Sec.~\ref{sec:3}, we first discuss the peculiarities of the EAC with respect to classical medium access control (MAC), and then we formalize the problem statement. In Sec.~\ref{sec:4}, we design the proposed EAC. In Sec.~\ref{sec:5}, we model and analyze the noise effects on the proposed EAC. In Sec.~\ref{sec:6}, we conduct a numerical analysis, aimed at providing guidelines and insights on the complex noise effects on the EAC through meaningful parameters. Finally in Sec.~\ref{sec:7}, we conclude the manuscript.

\section{Preliminaries}
\label{sec:2}

\subsection{System Model}
\label{sec:2.1}  

Multipartite entangled states constitute a key resource for implementing quantum information tasks \cite{HorHorHor-09}. When it comes to the generation of entangled states, it is very reasonable, given the current maturity of quantum technologies, to assume a specialized super-node responsible for the generation \cite{EppKamBru-17,AviRozWeh-22,VarGuhNai-21-1}. The rationale for this assumption is twofold. On one hand, it accounts for the  complex mechanisms and the dedicated equipment underlying the entanglement generation. On the other hand, it accounts for the mandatory requirement of some sort of local interaction among the qubits\footnote{With different entanglement classes characterized by different degrees of required qubit interactions. As instance, within graph states \cite{WalZweMus-16,HuaCheLi-23}, the required interactions are represented by the presence of edges in their graph representation.} to be entangled.\\

Accordingly, we consider a network in which a node, denoted in the following as $N_0$ and referred to as the \textit{orchestrator}, is in charge of the generation and distribution of the multiparty entangled state to be shared among the network nodes. The remaining $n$ nodes are denoted as $\{N_i\}_{i=1}^{n}$, and they contend for the multipartite entangled resource, referred to as \textit{``contended'' resource}. For this the following definition arises.

\begin{defin}[\textbf{Contention set}] 
    \label{def:01}
The \textit{contention set} $\mathcal{N}$ denotes the identities of the $n$ nodes:
    \begin{equation}
        \label{eq:01}
        \mathcal{N}=\{N_1, \ldots, N_n\}.
    \end{equation}
\end{defin}

When it comes to the distribution of the multipartite state, the orchestrator can, in principle, distribute each entangled qubit (ebit) to one node. However, this approach is not viable for all the classes of multipartite entanglement, which are characterized by different \textit{persistence} properties \cite{IllCalMan-22}. As an example, adopting this approach for distributing GHZ-like states, which are characterized by minimum persistence, implies the need that all the photons encoding the ebits are successfully distributed to the nodes in a single distribution attempt \cite{ZhoLiXia-23}. Alternatively, multipartite entangled states can be distributed through teleportation\cite{BugCouOma-23,AviRozWeh-22}, given a-prior distribution of EPR pairs via heralded scheme \cite{BarCroZei-10,HofHruOrt-12}. This strategy is practically ubiquitous and, indeed, in \cite{AviRozWeh-22} it has also been proved that such a strategy provides more resilience to noise and better protection against memory decoherence.  

Stemming from this, in the following, we adopt this approach for the entanglement distribution, which confers also generality to the analysis, being suitable for different classes of multipartite entanglement. Furthermore, the adopted heralded scheme allows the orchestrator to recognize which node -- if any -- experienced an ebit loss. In such a case, further distribution attempts can be performed to eventually distribute the targeted state to all the considered nodes.

\subsection{Definitions}
\label{sec:2.2}

 \begin{figure}
    \centering
    \begin{tikzpicture}[%
        every node/.style={
            font=\scriptsize,
            text height=1ex,
            text depth=.25ex,
        },
    ]
    \draw[->] (0.pt,0) -- (9,0);
    
    \draw[black,  thick] (8,12pt) -- (8,-4 pt);
    \draw[black,  thick] (2,12pt) -- (2,-4 pt);
    
    \draw[black,  thin] (0.6,3pt) -- (0.6,0 pt);
    \draw[black,  thin] (2.6,3pt) -- (2.6,0 pt);
    \draw[black,  thin] (3.6,3pt) -- (3.6,0 pt);
    \draw[black,  thin] (4.6,3pt) -- (4.6,0 pt);
    \draw[black,  thin] (5.6,3pt) -- (5.6,0 pt);
    \draw[black,  thin] (6.6,3pt) -- (6.6,0 pt);

    \node[anchor=north] at (2.3,0) {$\tau_g$};
    \node[anchor=north] at (3.2,0) {$\tau_d$};
    \node[anchor=north] at (4.2,0) {2$\tau_d$};
    \node[anchor=north] at (5.2,0) {$\cdots$};
    \node[anchor=north] at (6.2,0) {$M\tau_d$};
    \node[anchor=north] at (7.3,0) {$\tau_c$};
    \node[anchor=north] at (8.7,0) {t};
    
        node[anchor=north,midway,above=1pt] {Uplink};

    \draw[decorate,decoration={brace,amplitude=5pt}] (2,1) -- (8,1)
        node[anchor=south,midway,above=5pt] {$\tau_{th}$};     
    \end{tikzpicture}
    \caption{Simplistic representation of time-slotted sequence defined in Sec.\ref{sec:2.2} for the EAC.} 
    \label{fig:02}
    \hrulefill
\end{figure}
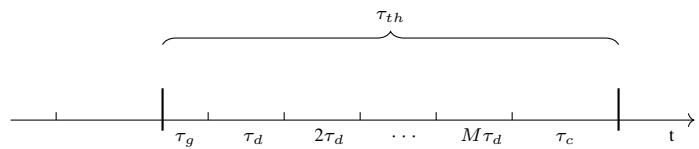

Here, we introduce some overall parameters which allow us to abstract from the particulars of the underlying quantum technology hardware as well as from the specificity of the considered multipartite state. This confers generality to the design and  analysis.

\begin{defin}[\textbf{Threshold Time}] 
    \label{def:02}
    The \textit{threshold time} $\tau_{th}$ denotes the maximum coherence time of the system, i.e., the time interval beyond which an entanglement resource would be irreversibly degraded by quantum decoherence.
\end{defin}

\noindent A multipartite state must be generated, distributed and utilized within this time interval.

\begin{defin}[\textbf{Multipartite Generation Time}]
    \label{def:03}
    The \textit{multipartite generation time} $\tau_{g}$ denotes the maximum time interval needed by the orchestrator for generating the targeted multipartite state.
\end{defin}

\begin{defin}[\textbf{Distribution Time}] 
    \label{def:04}
    The \textit{distribution time} $\tau_d$ denotes the maximum time interval needed by the orchestrator for one-attempt distribution of ebits to the considered nodes.
\end{defin}

\begin{rem}
    The distribution time depends on a multitude of factors, ranging from the characteristics of the (if present) matter-flying transducer \cite{CacCalTaf-20} through the quantum carrier/channel specificities to the individual link length. Nevertheless, we abstract from all these particulars through the notion of distribution time. Specifically, in case of noise-free entanglement distribution, $\tau_d$ models the all-inclusive time for generating the EPR pairs and distributing the corresponding ebits. Conversely, in case of noisy entanglement distribution, $\tau_d$ models likewise the time-interval needed for the EPR generation but it includes a single distribution attempt, as shown in Fig.~\ref{fig:02}.
\end{rem}

\begin{defin}[\textbf{Contention Time}]
    \label{def:05}
    The \textit{contention time} $\tau_c$ denotes the time interval needed for solving the entanglement contention problem, i.e., for accessing and utilizing the contended entangled state.
\end{defin}

\begin{defin}[\textbf{Successful Distribution Probability}]
    \label{def:06}
    The distribution attempts performed over a noisy quantum channel is modeled through a Bernoulli distribution with parameter $p$, with $p$ denoting the probability of successfully distributing an ebit to one of the end-nodes.
\end{defin}

\noindent From Def.~\ref{def:06}, we consider, for the EPR distribution process, the absorbing quantum channel model\footnote{ It is worthwhile to mention that when it comes to free-space channels the transmission conditions results as more complex and particularly adverse \cite{Djo-20,Djo-21,NafDjo-23}. Nevertheless, the model here proposed represents the \textit{worst-case scenario}, since the noise irreversibly corrupts the information carrier without any possibility of further ebit recovery.} \cite{NieChu-11, BenDiVSmo-97,BenShoSmoTha-99,BruFaoMac-00}. This channel model is characterized by two elementary events: i) \textit{E= "successful transmission"}, and ii) the corresponding complementary event $\Bar{E}=$\textit{"failed transmission"}, representing the loss, i.e., the absorption, of the transmitted particle encoding the ebit.

\begin{rem}
    
    The adopted quantum channel model can be easily extended to account for the fidelity of the distributed EPR pairs.  
    Indeed, ebits received with low fidelity impact on the fidelity of the teleported state \cite{Joz-94,NieChu-11} and hence on the performance of the EAC protocol via the fidelity of either the contention resolution state or the contended resource state. Thus, whenever the fidelity of the distributed EPR pairs $F$ may results below a certain fidelity threshold $F_{th}$, the distribution attempt can be considered as failed, as it does not meet the requirements of the EAC protocol. In fact, this event prevents the correct teleportation of the multipartite entangled state. In this light, the events \textit{E= "successful transmission"} and $\Bar{E}=$\textit{ "failed transmission"} correspond to the events \textit{"ebit distributed with fidelity $F>F_{th}$"} and \textit{"ebit distributed with fidelity $F<F_{th}$"}, respectively. Clearly, the value of $F_{th}$ depends on the particulars of the application exploiting the EAC protocol. On the other hand, when the fidelity of the entanglement distribution is above the given threshold, $F>F_{th}$, one may still wish to perform entanglement purification to improve the fidelity of the distributed EPRs and, hence, of the teleported states. In such a case, each node would require more than one ebit distribution, as we will analyze in Sec.~\ref{sec:5}.
\end{rem}

\begin{defin}[\textbf{Time Horizon}]
    \label{def:07}
    The time horizon $M$ denotes the maximum number of distribution attempts that can be performed within the threshold time interval $\tau_{th}$:
    \begin{equation}
        \label{eq:02}
        M = \left\lfloor \left[ \tau_{th} - \left( \tau_g + \tau_c \right)\right] / \tau_d \right\rfloor
    \end{equation}
\end{defin} 

\noindent In \eqref{eq:02}, 
we consider the worst case scenario, since we evaluate $M$ by assuming that each $\tau_d$ includes the EPR generation time. In the case of an orchestrator with unconstrained number of communication qubits \cite{CalAmoFer-22}, this restriction can be removed by generating the EPRs for different distribution attempts in parallel. 

\begin{defin}[\textbf{Connected set}]
    \label{def:08}
    The connected set $\mathcal{S}_m^j\subseteq  \mathcal{N}$ denotes the set constituted by $j \leq n$ nodes for which a successful transmission of an ebit has been accomplished within the $m$-th distribution attempt, with $m \in \{1,2, \ldots, M\}$.
\end{defin}
 As instance, $\mathcal{S}_2^3$ denotes the connected set at slot $m=2$ in which 3 nodes correctly received the ebits. 
\noindent In the ideal scenario of noiseless EPR distribution, $\mathcal{S}_m^j=\mathcal{S}_m^n \equiv \mathcal{N}$ with probability one for $m=1$, namely, after a single distribution attempt of duration $\tau_d$. Conversely, in presence of noisy quantum channels, it generally results $\mathcal{S}_m^j \subseteq \mathcal{N}$, with $j \leq n $ and there exists a no-null probability of having $\mathcal{S}_M^j \subset \mathcal{N}$, with $j \neq n$ at the threshold time $\tau_{th}$, as we will show in the next sections. Furthermore, by accounting for the developed system model, it results:
\begin{equation}
    \label{eq:03}
    \mathcal{S}_{m-1}^j \subseteq \mathcal{S}_{m}^i, \, \text{with} \, j \leq i, \, \forall \, m \in \{2, \ldots, M\}.
\end{equation}


\section{EAC: Entanglement Access Control}
\label{sec:3}

In this section, we first discuss the peculiarities of  \textit{entanglement access control} (EAC) with respect to classical medium access control (MAC), and then we formalize the problem statement.

\subsection{MAC vs EAC}
\label{sec:3.1}

An EAC protocol aims at solving the contention problem arising when an entangled resource is shared among multiple networks nodes. From this definition, the design of an EAC protocol is reminiscent of the design of medium access control (MAC) protocols in classical networks \cite{KurRos-12}. Formally, the overall goal of a MAC protocol is to determine the functional $\chi(\cdot)$ univocally selecting the network node granted with the access to a communication resource\footnote{Depending on the adopted access strategy, the communication resource could be, as instance, a certain time-slot or a certain frequency band (in deterministic access protocols), or perhaps the utilization of the channel in random access strategies.}:
\begin{align}
    \label{eq:04}
    \chi :& \, \mathcal{N} \rightarrow \{0,1\}, 
    & \textrm{ s.t. } \exists! N_i \in \mathcal{N} : \chi(N_i) = 1
\end{align}

Conversely, the \textit{non-local scope} and the \textit{global, dynamic utility} exhibited by entanglement complicate the problem, as pointed out in \cite{IllCalMan-22}.
First, there must be a tight cooperation between the network nodes – nodes that must be aware of each other identities – storing the entangled qubits for being able to exploit the quantum correlation provided by entanglement. Furthermore, any processing of a single entangled qubit has an instantaneous effect on the global entangled state, with possible changes affecting the remaining entangled qubits as well, regardless of the distances among the entangled nodes. These network entities even compete among each others to use the same entanglement resource. As instance with reference to the teleportation protocol \cite{CacCalVan-20}, any node sharing entanglement can act either as source or as destination, as long as it coordinates with the other entangled nodes. Hence, when it comes to the design of the network functionalities, access to the entanglement resource is pivotal.

Furthermore, multipartite entanglement enriches the connectivity features. Specifically, multipartite entangled states allow dynamic selection -- rather than fixed a-priory node selection, during the entanglement generation and distribution process stage -- of the nodes granted with the right to utilize the entanglement resource. And this dynamic selection can be performed according to the communication needs \cite{LiNie-13,DuaZhaSun-14} or according the specific application, spanning from computing \cite{CalAmoFer-22}, through to quantum secret sharing \cite{ZhaLiMan-05} to clock synchronization \cite{KomKesBis-14}. In this light, by accounting for the no-broadcasting theorem -- which prevents from broadcasting an unknown quantum state to two or more receivers -- multipartite entanglement seems reminiscent of multi-point channels, but in the broad sense of allowing distributed computing and communication tasks. From the above discussion, it follows that the overall goal of the EAC protocol is to determine the functional $\chi(\cdot)$ univocally selecting the \textit{subset} of $k\leq n$ network nodes granted with the right to utilize the multiparty entangled resource, for fulfilling a certain computing or communication task. Formally:

\begin{align}
    \label{eq:05}
    \chi : \mathcal{N}^k\rightarrow \{0,1\}& 
     \nonumber\\\textrm{ s.t.    } \exists! &\text{ } k\text{\rm -tuple } (N_{i_1},N_{i_2},\ldots, N_{i_k}) \in \mathcal{N}^k:\nonumber\\
     &\chi(N_{i_1},N_{i_2},\ldots, N_{i_k}) = 1.
\end{align}
The simplest problem the EAC can solve is the one with $k=2$, i.e., a couple of nodes in $\mathcal{N} \times \mathcal{N}$. 

\subsection{Problem Statement}
\label{sec:3.2}

Accordingly, the EAC design is conducted by accounting for these key requirements:
\begin{itemize}
    \item[-] The protocol must be able to distributively solve the contention problem by \textit{univocally} determining the subset of $k$ nodes (among $n$ possible candidates) granted with the access to -- namely, the right to utilize -- the entangled contended resource. 
    \item[-] The protocol must support the \textit{anonymity} of the selected nodes, i.e., their identities are kept hidden to each others as well as to the un-selected nodes. Conversely, this information about the selected node identities is made available at a trusted node -- represented by the orchestrator -- to be eventually exploited for implementing further network functionalities\footnote{Indeed, controlling the access of network nodes to entanglement as resource may be a crucial functionality in many communication scenarios.  And, generally, controlling the access to a shared resource also entails retaining information on the identities of the nodes accessing the resource. From this perspective, it is reasonable to assume that this information is available at a dedicated node, namely, the orchestrator, delegated for maintaining track of the evolution of the resource accesses.}.
    \item[-] The protocol must abstain from delegating the contention solution to the classical network, which, in turn, would require a functional -- but, still, a research open issue -- classical-quantum interface between the classical Internet and the Quantum Internet.
\end{itemize}

We note that the joint requirements of \textit{distributed strategy}, \textit{anonymity} and \textit{no classical signaling} impose another constraint on the EAC with respect to classical MAC protocols: the contention solution must be disclosed at each node as a result of a local processing, as analyzed in Sec.~\ref{sec:4}.

\section{Quantum-Genuine EAC Design}
\label{sec:4}
Here, we design a ``quantum-genuine'' EAC, which fulfils the requirements in Sec.~\ref{sec:3.2}, by relying on the features of a singled-out multipartite entangled state, referred to as ``\textit{contention-resolution state}''. By exploiting its features, we deterministically solve the contention problem arising in accessing to an entangled resource, referred to as ``\textit{contended}'' resource in the following.

\subsection{Multipartite Entanglement Resource for the Contention Resolution}
\label{sec:4.1}
\begin{figure*}[t]
	\centering
	\begin{minipage}[c]{.48\linewidth}
		\begin{center}
          \begin{adjustbox}{width=0.6\textwidth}
               \begin{quantikz}
                    & & \lstick[wires=4]{$\ket{D_4^2}$} 
                         &\qw\gategroup[wires=7,steps=3,style={dashed,rounded corners,inner xsep=5pt,inner ysep=5pt}, background, label style={label position=above, yshift=-0.0cm}, background]{\sc Encoder} &\qw &\ctrl{4} &\qw&\qw\rstick[wires=1]{$D_1$} & & \rstick[wires=7]{$\ket{\Lambda_7}$}\\
                    & & &\qw &\ctrl{4} &\qw &\qw&\qw\rstick[wires=1]{$D_2$}& &\\
                    & & &\ctrl{4} &\qw &\qw &\qw&\qw\rstick[wires=1]{$D_3$}& &\\
                    & & &\qw &\qw &\qw &\qw &\qw\rstick[wires=1]{$D_4$}& &\\
                    & & \lstick{$\ket{0}$} & \qw &\qw &\targ{} &\qw &\qw\rstick[wires=1]{$A_0$} & &\\
                    & & \lstick{$\ket{0}$} & \qw &\targ{} &\qw &\qw &\qw\rstick[wires=1]{$A_1$}& &\\
                    & & \lstick{$\ket{0}$} & \targ{} &\qw &\qw &\qw &\qw\rstick[wires=1]{$A_2$}& &
                \end{quantikz}
          \end{adjustbox}
    \end{center}
		\subcaption{\footnotesize{Quantum circuit corresponding to a \textit{linear} contention-resolution encoder for $n = 4$ and $k=2$. The linear encoder requires a number $\ell=n-1$ of ancillary qubits.}}
  
		\label{fig:03a}
	\end{minipage}
    \hfill
	\begin{minipage}[c]{.48\linewidth}
    \centering

    \begin{adjustbox}{width=0.48\textwidth}
        \begin{tabular}{ c | c | c | c || c | c | c }
            \toprule
            $d_1$ & $d_2$ & $d_3$ & $d_4$ & $a_0$ & $a_1$ & $a_2$  \\
            \midrule
            1 & 1 & 0 & 0 &   1 & 1 & 0 \\
            1 & 0 & 1 & 0 &   1 & 0 & 1 \\
            0 & 1 & 1 & 0 &   0 & 1 & 1 \\
            1 & 0 & 0 & 1 &   1 & 0 & 0 \\
            0 & 1 & 0 & 1 &   0 & 1 & 0 \\
            0 & 0 & 1 & 1 &   0 & 0 & 1 \\
            \bottomrule
        \end{tabular}
    \end{adjustbox}

    \subcaption{\footnotesize{Measurement outputs for the \textit{linear} encoder represented in Fig.~\ref{fig:03a}. Each 
    ancilla measurement outcome is mapped to only one of the $\binom{n}{k} = 6$ possible node selection configurations. }}
    
    \label{tab:02}
	\end{minipage}
	\caption{Linear contention-resolution encoder}
	\label{fig:03}
    \hrulefill
\end{figure*}
In this manuscript, we propose to solve the contention problem described in Sec.~\ref{sec:3} by exploiting the features of a particular class of multipartite entangled states, namely, the \textit{Dicke states} \cite{Dic-54,ManWol-95,BarEid-22}, which can be deterministically generated at a node as proven in \cite{BarEid-22}.

A n-qubit Dicke state, denoted as $\ket{D^k_n}$, is an even superposition of the $n$-qubit computational basis states ${\ket{s}}_{s \in \{0,1\}^n}$, with each state characterized by a Hamming distance $d_H(\cdot)$ equal to $k$:
\begin{equation}
    \label{eq:06}
    \ket{D^k_n} =\left[ \binom{n}{k}\right]^{-\frac{1}{2}} \sum_{s\in\{0,1\}^n : d_H(s) = k}\ket{s}.
\end{equation}

For designing the quantum-genuine EAC protocol granting entanglement access to a subset of $k$ among $n$ nodes, we propose to exploit the $n$-qubit Dicke state $\ket{D_n^k}$ and, in particular, to set the value of $k$ in \eqref{eq:06} as the cardinality of the subset of $n$ network nodes granted with the right to utilize the contended entangled resource. 
More into detail, we consider the state $\ket{\Lambda_{n+\ell}}$, referred to as \textit{contention-resolution} state. This state is generated at the orchestrator by enriching a $\ket{D_n^k}$ state with $\ell$ ancillary qubits processed through a proper encoder, described in Sec.~\ref{sec:4.2}. Formally, we have:
\begin{equation}
    \label{eq:07}
    \ket{\Lambda_{n+\ell}}= \ket{D_1,\cdots,D_n,A_0,\cdots,A_{\ell-1}},
\end{equation}
with $D_i$ denoting the $i$-th qubit of the Dicke state and $A_j$ the $j$-th ancillary qubit processed by the encoder.\\

Once the contention-resolution state $\ket{\Lambda_{n+\ell}}$ has been generated at the orchestrator, it is distributed among the nodes within the contention set $\mathcal{N}$ through teleportation, as motivated in Sec.~\ref{sec:2.1}. Indeed, the orchestrator retains at its side the ancillary qubits $A_0, \ldots, A_{\ell-1}$, while teleporting the $i$-th qubit of the Dicke state $\ket{D^k_n}$ to node $N_i$. As described in the following, the rationale for retaining the ancilla qubits at the orchestrator is to provide the orchestrator with the identities of the nodes selected from the EAC, without the need of exchanging classical signaling and, thus, by preserving their anonymity. This is a key feature of the proposed quantum-genuine EAC, since it allows the possibility to exploit such an information later for requesting network functionalities, as exemplified within Sec.~\ref{sec:4.3}. Once the contention-resolution state has been successfully distributed, the EAC protocol works as follows. Each node $N_i \in \mathcal{N}$ performs a local measurement (in the computational basis) of the $i$-th qubit of the contention-resolution state $\ket{\Lambda_{n+\ell}}$ available at its side. By denoting with $d_i \in \{0,1\}$ the measurement outcome, whenever $d_i=0$, node $N_i$ becomes aware it lost the contention. Otherwise, whenever $d_i=1$, node $N_i$ is granted access to the contended resource, i.e., it gains the right to exploit the entanglement resource. 
By accounting for \eqref{eq:06}, it results that exactly $k$ nodes of the contention set $\mathcal{N}$ observe the measurement outcome $1$, whereas the remaining $n-k$ nodes observe the outcome $0$. Since each state -- representing one of the possible $\binom{n}{k}$ subsets of $k$-selected nodes -- in the superposition \eqref{eq:06} exhibits the same amplitude, each state experiences the same probability of being observed. And this probability is equal to $1/\binom{n}{k}$. Furthermore, the states in \eqref{eq:06} associated with a positive measurement outcome for a certain node $N_i$ (i.e., $d_i = 1$) differ each other for a permutation of $k-1$ ``1s'' over $n-1$ positions. And there exist $\binom{n-1}{k-1}$ of such possible permutations. As a consequence, each node in $\mathcal{N}$ experiences the same probability of being granted with the access to the \textit{contended resource} equal to $\binom{n-1}{k-1}/\binom{n}{k}=k/n$\footnote{As an example, the Dicke state $\ket{D^2_4}$ enables a fair selection of a subset of 2 out of 4 nodes with probability $1/6$. And each individual node experiences a probability of being granted with access to the contended resource equal to $1/2$.}.
From the above, due to the Dicke state features, the proposed EAC allows a fair -- namely, equiprobable -- resolution of the contention problem among $n$ nodes, by distributively enabling $k$ nodes to access the contended state, singled out accordingly to the considered communication/computing task. Clearly the value of $k$ depends on the specific network application and, hence, on the size of the cluster to be determined. 

\subsection{Contention-Resolution Encoder}
\label{sec:4.2}

As aforementioned, our proposal exploits the contention-resolution state $\ket{\Lambda_{n+\ell}}$, generated at the orchestrator by processing $\ell$ ancilla qubits with a suitable encoder.\\

The aim of such an encoder, referred to as \textit{contention-resolution encoder}, is to process the ancillary qubits, without any alteration of the initial Dicke state $\ket{D^k_n}$, so that each possible measurement outcome of the ancillas is one-to-one mapped to only one of the $\binom{n}{k}$ possible configuration in \eqref{eq:06}, which in turn represents a different subsets of $k$ nodes among $n$ granted with the access to the contended resource. \\

To this aim, the contention-resolution encoder is designed to enforce an orthogonality condition on the ancilla quantum states. In this way, the orchestrator, by measuring the ancilla qubits, is able to discriminate the node subset (among the $\binom{n}{k}$ different subsets) determined by running the proposed genuine EAC, without exchanging classical signaling nor publicly revealing the node identities. It must be noted, though, that there exist multiple choices to design the contention-resolution encoder by satisfying the aforementioned requirements. In the following, we discuss some of these choices by focusing on two fundamental parameters, which play a key role in the encoder complexity, namely, the number of ancillary qubits and the number of multi-qubit  gates\cite{XuTon-22}. Specifically, we provide some insights and guidelines, by optimizing the design with respect to one of the two aforementioned key parameters, and by discussing the impact of this optimization on the second parameter.
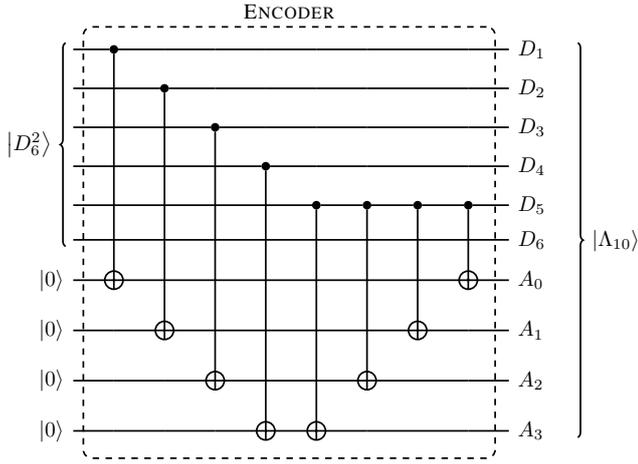
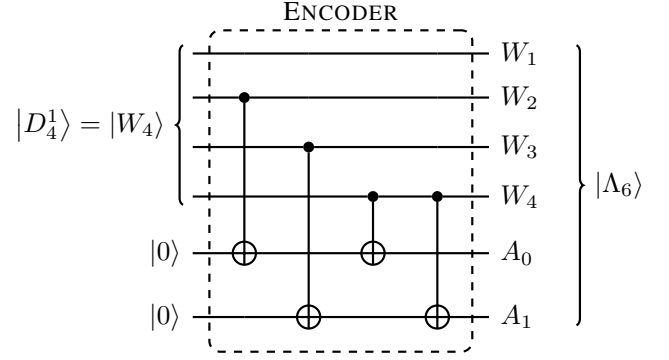
\begin{figure*}[t]
	\centering
	\begin{minipage}[c]{.47\linewidth}
		\begin{center}
        \begin{adjustbox}{width=\textwidth}
            \begin{quantikz}
                & & \lstick[wires=6]{$\ket{D_6^2}$} &\ctrl{6} &\gategroup[wires=10,steps=8,style={dashed,rounded corners,inner xsep=10pt,inner ysep=5pt,xshift = -0.7cm}, background, label style={label position=above, xshift = 0.0cm, yshift=-0.0cm}, background]{\sc Encoder}\qw &\qw &\qw&\qw&\qw&\qw&\qw&\qw\rstick{$D_1$}& &\rstick[wires=10]{$\ket{\Lambda_{10}}$}\\  
                & & &\qw &\ctrl{6} &\qw &\qw &\qw& \qw &\qw&\qw&\qw\rstick{$D_2$}& &\\
                & & &\qw &\qw &\ctrl{6} &\qw &\qw &\qw&\qw&\qw&\qw\rstick{$D_3$}& &\\
                & & &\qw &\qw &\qw &\ctrl{6} &\qw &\qw &\qw &\qw&\qw\rstick{$D_4$}& &\\
                & & &\qw &\qw &\qw &\qw &\ctrl{5}& \ctrl{4} &\ctrl{3}&\ctrl{2}&\qw\rstick{$D_5$}& &\\
                & & &\qw &\qw &\qw &\qw &\qw &\qw &\qw &\qw&\qw\rstick{$D_6$}& &\\
                & & \lstick{$\ket{0}$} & \targ{} &\qw &\qw &\qw&\qw &\qw &\qw &\targ{}&\qw\rstick{$A_0$}& &\\
                & & \lstick{$\ket{0}$} & \qw &\targ{} &\qw&\qw &\qw &\qw &\targ{} &\qw &\qw\rstick{$A_1$}& &\\
                & & \lstick{$\ket{0}$} & \qw &\qw &\targ{} &\qw&\qw&\targ{}&\qw &\qw&\qw\rstick{$A_2$}& &\\
                & & \lstick{$\ket{0}$} & \qw &\qw &\qw &\targ{}&\targ{}&\qw&\qw&\qw&\qw\rstick{$A_3$} & &
            \end{quantikz}
        \end{adjustbox}
    \end{center}
		\subcaption{\footnotesize{Quantum circuit corresponding to the \textit{binary} contention-resolution encoder for $n = 6$ and $k=2$. The encoder requires a number $\ell = \lceil \log_2{\binom{n}{k}}\rceil = 4$ of ancillary qubits.}}
  
		\label{fig:04a}
	\end{minipage}
    \hfill
	\begin{minipage}[c]{.47\linewidth}
    \centering
    \begin{center}
         \begin{adjustbox}{width=\textwidth}
            \begin{quantikz}
                \lstick[wires=4]{$\ket{D^1_4}=\ket{W_4}$} & \gategroup[wires=6,steps=4,style={dashed,rounded corners,inner xsep=5pt,inner ysep=5pt}, background, label style={label position=above, yshift=-0.0cm}, background]{\sc Encoder} \qw & \qw & \qw & \qw & \qw\rstick{$W_1$}  & &\rstick[wires=6]{$\ket{\Lambda_6}$} \\
                 & \ctrl{3} & \qw & \qw & \qw & \qw\rstick{$W_2$} & & \\
                 & \qw & \ctrl{3} & \qw & \qw & \qw\rstick{$W_3$} & & \\
                 & \qw & \qw & \ctrl{1} & \ctrl{2} & \qw\rstick{$W_4$} & & \\
                 \lstick{$\ket{0}$} & \targ{}& \qw & \targ{} & \qw & \qw\rstick{$A_0$} & &  \\
                 \lstick{$\ket{0}$} & \qw & \targ{} & \qw & \targ{} & \qw\rstick{$A_1$} & & 
            \end{quantikz}
        \end{adjustbox}
    \end{center}
    \subcaption{\footnotesize{Quantum circuit corresponding to the \textit{binary} encoder for $n = 4$ and $k=1$. The encoder requires a number $\ell =\lceil \log_2{n}\rceil =2$ of ancillary qubits as specified by \eqref{eq:09}.}}
    \label{fig:04b}
	\end{minipage}
	\caption{Binary contention-resolution encoder.}
	\label{fig:04}
    \hrulefill
\end{figure*}

\subsubsection{Linear Encoder}
Stemming from the above, a possible  choice is constituted by the encoder represented in Fig.~\ref{fig:03} and referred to as \textit{linear encoder}.  This encoder requires a number $\ell$ of ancillary qubits equal to $\ell=n-1$, and each ancillary qubit is elaborated by a CNOT gate controlled by a different qubit of the Dicke state. Within the figure and without any loss of generality, the controlled gate $\texttt{CNOT}(D_{i+1},A_i)$ acting on the $i$-th ancilla is controlled by the $(i+1)$-th qubit of the Dicke state. The different configurations\footnote{In the following, we will widely use a sequential enumeration of binary strings,  assuming \textit{big-endian} endianness.} that are obtained through the ancilla measurement are reported in Fig.~\ref{tab:02}. Each of these ancilla configurations is associated to a different subset of $k$ nodes granted with the access to the contended resource.
Indeed, it is easy to verify that the linear encoder ensures the orthogonality condition on the ancillary qubits. In fact, let us suppose, without any loss in generality, that the qubit labeled as $D_{n}$ is the one not controlling any CNOT operation (as represented in Fig.~\ref{fig:03}). Furthermore, let us consider two arbitrary states -- say $\ket{s}$ and $\ket{s'}$ -- in \eqref{eq:06}. The overall action of the CNOTs within the linear encoder is to map the controlling qubit configurations on the ancillary qubits, so that whenever the control is measured as zero, the ancilla is measured as zero as well, and vice-versa. 
Clearly, if the $n$-th qubits of $\ket{s}$ and $\ket{s'}$ differ, then we have from \eqref{eq:06} that the two hamming distances computed on the first $n-1$ qubits differ as well. This along with the structure of Dicke states in \eqref{eq:06} imply that the first $n-1$ qubit configurations  can be regarded as orthogonal states in a $n-1$-qubit system. And this orthogonality is directly mapped by the encoder into the ancillary qubits. \\

Conversely, whenever the $n$-th qubits of $\ket{s}$ and $\ket{s'}$ are the same, the hamming distances computed on the first $n-1$ qubits are the same. Yet, given that the Dicke state in \eqref{eq:06} is a superposition of different $n$-qubits computational basis states, $\ket{s}$ and $\ket{s'}$ must differ in at least two of the first $n-1$ qubits. This, once again, implies that the first $n-1$ qubit configurations can be regarded as orthogonal states in a $n-1$-qubit system. And this orthogonality is directly mapped into the ancillary qubits. 
It is straightforward to recognize that the information about the $n$-th qubit of the Dicke state $\ket{D^k_n}$ can be directly recovered through a ``parity'' check over the $n-1$ ancillary qubits. Specifically, by denoting with $a_i$ the outcome of the measurement operation performed over the $i$-th ancillary qubit, whenever $\sum_{i=0}^{n-1} a_i = k$, then the measurement outcome $d_n$ corresponding to the $n$-th
qubit of the Dicke state is $d_n=0$. Conversely, whenever  $\sum_{i=0}^{n-1} a_i = k-1$, then $d_n=1$. The linear encoder requires $n-1$ ancillary qubits as well $n-1$ CNOT operations and, hence, its complexity in terms of gate count \cite{FerCacAmo-21,CalAmoFer-22} scales linearly with the number of nodes competing for the entanglement resource. Yet, it is possible to conceive encoders with a smaller number of ancillary qubits, at the price of an higher number of CNOT operations, as detailed in the following. 

\subsubsection{Binary Encoder} Specifically, it is possible to minimize the number of ancillary qubits by considering a different choice for the encoder, referred to as \textit{binary encoder}. This encoder univocally maps the subscripts $\{i\}$ of the $\binom{n}{k}$ orthogonal states $\{\ket{s_i}\}$ in \eqref{eq:06} into binary-coded decimal words. Thus, each $\ket{s_i}$, through its subscript, has a univocal associated binary-coded decimal representation, and this implies a number of ancillary qubits equal to $l = \lceil \log_2 \binom{n}{k} \rceil$.

\begin{figure}[t]
    \centering
    \includegraphics[width=1\columnwidth]{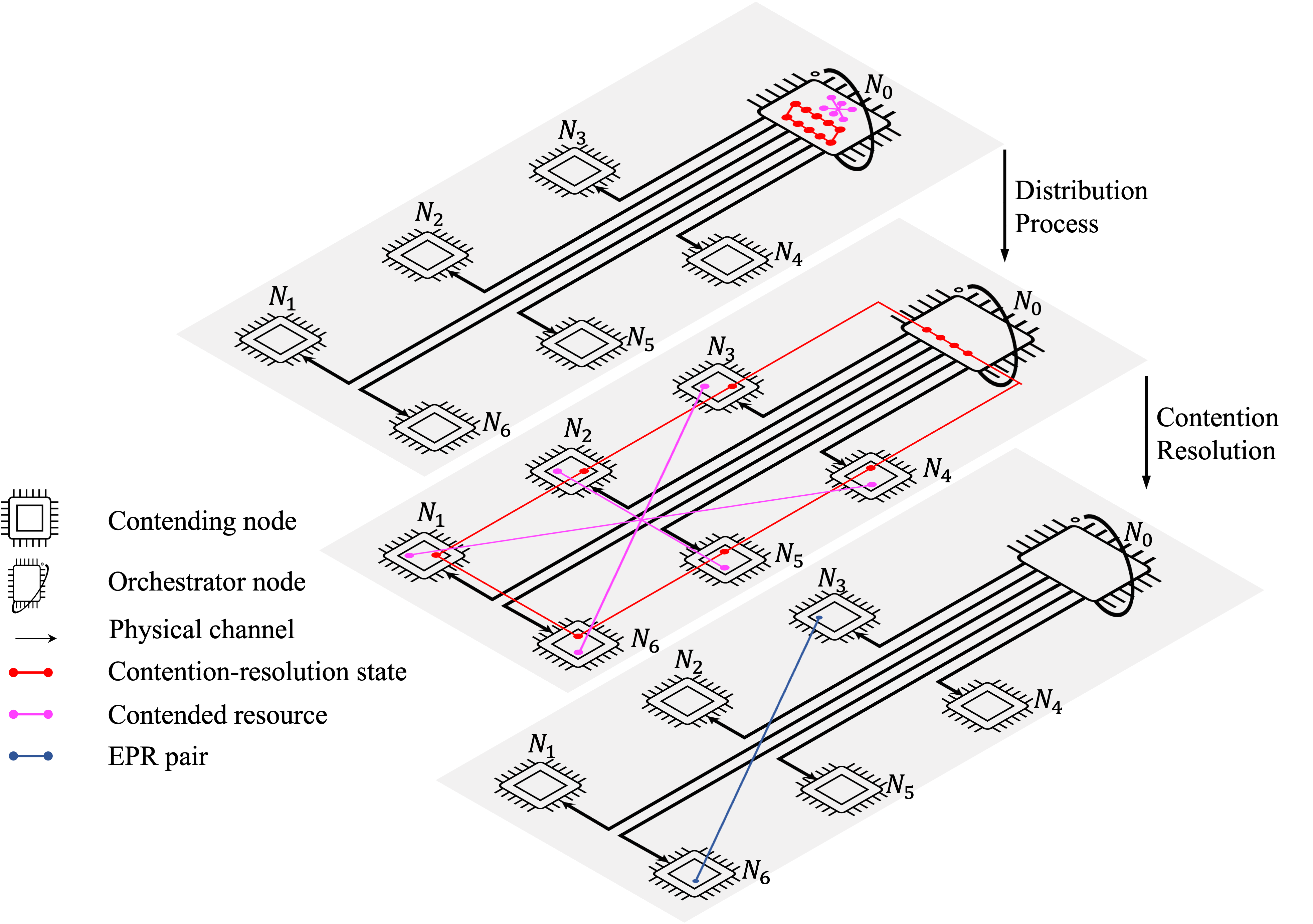}
    \caption{Graphical representation of the EAC protocol.}
    \label{fig:06}
    \hrulefill
\end{figure}

As regards to the number of required CNOTs required by the binary encoder, there exist different strategies depending on the particular setting for $n$ and $k$. One of these strategy consists in applying a sequence of CNOTs accordingly to the following rule. We map the arbitrary state in the superposition \eqref{eq:06} with subscript $i+1$ to the binary representation of $i$:
\begin{equation}
    \label{eq:08}
    i = \sum_{j=0}^{\lceil \log_2{\binom{n}{k}}\rceil-1 } b^i_j 2^j
\end{equation}
The considered mapping requires controlled operations with the qubits of the Dicke state acting as controls and the ancillary qubits acting as targets. An example of such a binary encoder for $\ket{D_6^2}$ is given in Fig.~\ref{fig:04a}. The considered example for a \textit{binary} encoder requires only $8$ CNOTs. This is achieved by wisely exploiting the presence of multiple qubits in $\ket{1}$ state within each state of the superposition in \eqref{eq:06}, for achieving univocal configurations through the ancillary qubits measurements.

Remarkably, whenever $k=1$ the Dicke state $\ket{D^k_n}$ reduces to a W-state $\ket{W_n}$. In such a case, the aforementioned rule can be specialized as follows. Specifically, we map the arbitrary state in \eqref{eq:06} where the $i+1$-th qubit is $\ket{1}$ to the binary representation of $i$:
\begin{equation}
    \label{eq:09}
    i = \sum_{j=0}^{\lceil \log_2 i-1 \rceil} b^i_j 2^j
\end{equation}
The considered mapping requires a $\texttt{CNOT}(W_{i+1},A_{j})$ -- with the $(i+1)$-th qubit of $\ket{W_n}$ acting as control and the $j$-th ancillary qubit acting as target --  for any $b^i_j \neq 0$ in \eqref{eq:09}. Such an encoder is represented in Fig.~\ref{fig:04b} for a particular case, namely, for $n=4$ contending nodes. Accordingly, the number of CNOT operations required by the proposed encoder is:
\begin{equation}
    \label{eq:10}
    \sum_{j=1}^\ell \binom{\ell}{j} j = \ell \,2^{\ell-1} = \lceil\log_2 n\rceil \, 2^{\lceil\log_2 n\rceil-1}
\end{equation}
\eqref{eq:10} is exactly the number of required CNOT operations when $n$ is power of $2$. Conversely, whenever $n$ is not power of $2$, \eqref{eq:10} overestimates the number of CNOT operations, due to the ceil-function determining the number of ancillary qubits.
Indeed, the number of CNOTs in \eqref{eq:10} represents an upper bound of the required number of gates whenever $k>1$, regardless whether $n$ is a power of $2$ or not, as shown with the exemplary encoder represented 
in Fig.~\ref{fig:04a}.

\subsection{Genuine EAC: Example for $k=2$}
\label{sec:4.3}
As representative example of the genuine EAC protocol, we consider the task of fairly selecting $k=2$ out of $n$ nodes for accessing to the contended entanglement resource, as depicted in Fig.~\ref{fig:06}.
This setting models, as instance, a scenario where the nodes compete each other to extract, from a GHZ state, an EPR pair at run-time to be subsequently used for a certain communication/computing task.
More into details, an $n$-qubit GHZ state \cite{GreHorZei-89} acts as contended resource. The maximally connectedness property \cite{NieChu-11} of an ideal GHZ state allows the extraction of an EPR pair that is: i) \textit{deterministic}; ii) \textit{invariant with respect to node identities}, i.e., an EPR pair can be extracted between any pair of nodes sharing the GHZ state; iii) \textit{LOCC}, i.e., the extraction relies on local operations and classical communications.
By adopting the proposed genuine EAC, the contention resolution works as follows. Each contending node locally holds one qubit of the $n$-qubits GHZ state. The orchestrator generates and distributes the contention-resolution state $\ket{\Lambda_{n+\ell}}$ obtained by exploiting the Dicke state $\ket{D^2_n}$ as detailed above. Then, the competing nodes perform a measurement on the qubits of $\ket{\Lambda_{n+\ell}}$ at their side, by obtaining the corresponding measurement outcomes $d_1,\cdots,d_n$. Such measurement results substitute the classical control signaling, and they act as logical control for the operation to be performed on the contended GHZ state.
More into details, whenever $d_i=0$ -- i.e., whenever $N_i$ \textit{loses} the contention -- $N_i$ performs a local measurement in the Hadamard basis on the $i$-th qubit (namely, the qubit at its side) of the contended resource $\ket{GHZ_n}$, with measurement outcome denoted as $g_i\in\{0,1\}$. This measurement guarantees $N_i$ ``to safely leave'' the entangled state $\ket{GHZ_{n}}$, by preserving the entanglement shared between the remaining nodes.
Conversely, whenever $d_i=1$, $N_i$ is granted access to the contended resource. As already mentioned, according to \eqref{eq:06}, exactly $n-k$ nodes will lose the contention while the remaining $k=2$ nodes are allowed to access the resource.
The value of $d_i$ determines the local unitary to be performed on the $i$-th GHZ qubit at the $i$-th node for distributively extracting an EPR pair, as follows:
\begin{align}
    \label{eq:12}
    U_{d_i} &= \begin{cases}
        H & \textrm{if } d_i=0\\
        I & \textrm{if } d_i=1\\
    \end{cases}
\end{align}
with $I$ and $H$ denoting the identity and the Hadamard unitary, respectively.
Let $\textbf{d}$ denote the ordered vector of the contention resolution measurement results, namely, $\textbf{d}=[d_1,\cdots,d_n]$ with the index $i$ denoting the $i$-th node identity.
From \eqref{eq:12}, it results that the overall unitary $U_{\textbf{d}}$ acting on the $n$-GHZ state is given by: $U_{\textbf{d}} = U_{d_1} \otimes \ldots \otimes U_{d_n}$.

Let us assume, without loss of generality, the indexes $i$ and $j$ corresponding to the identities of the contention winner nodes. By applying the operator $U_\textbf{d}$ to the GHZ state, after some algebraic manipulations, the output state is: 
\begin{align}
    \label{eq:14}
    U_\textbf{d}\ket{GHZ_{n}} = & \ket{\Phi^+} \otimes \sum_{\substack{\psi\in\{0,1\}^{n-2}\\d_H(\psi) \text{ even}}} \ket{\psi} + \ket{\Phi^-} \otimes \sum_{\substack{\psi'\in\{0,1\}^{n-2}\\d_H(\psi') \text{ odd}}} \ket{\psi'}
\end{align}
In \eqref{eq:14}, $\ket{\Phi^+}$ and $\ket{\Phi^-}$ denote the two Bell states resulting from the actions of the two identities $i$ and $j$. Instead, the $(n-2)$-qubit states $\ket{\psi}$ and $\ket{\psi'}$ denote the states resulting from the processing induced by the $(n-2)$ unitaries $H$. These states are characterized by an \textit{even} and \textit{odd} number of qubits in state $\ket{1}$, respectively.

\begin{figure}[t]
    \centering
    \includegraphics[width = 1\columnwidth]{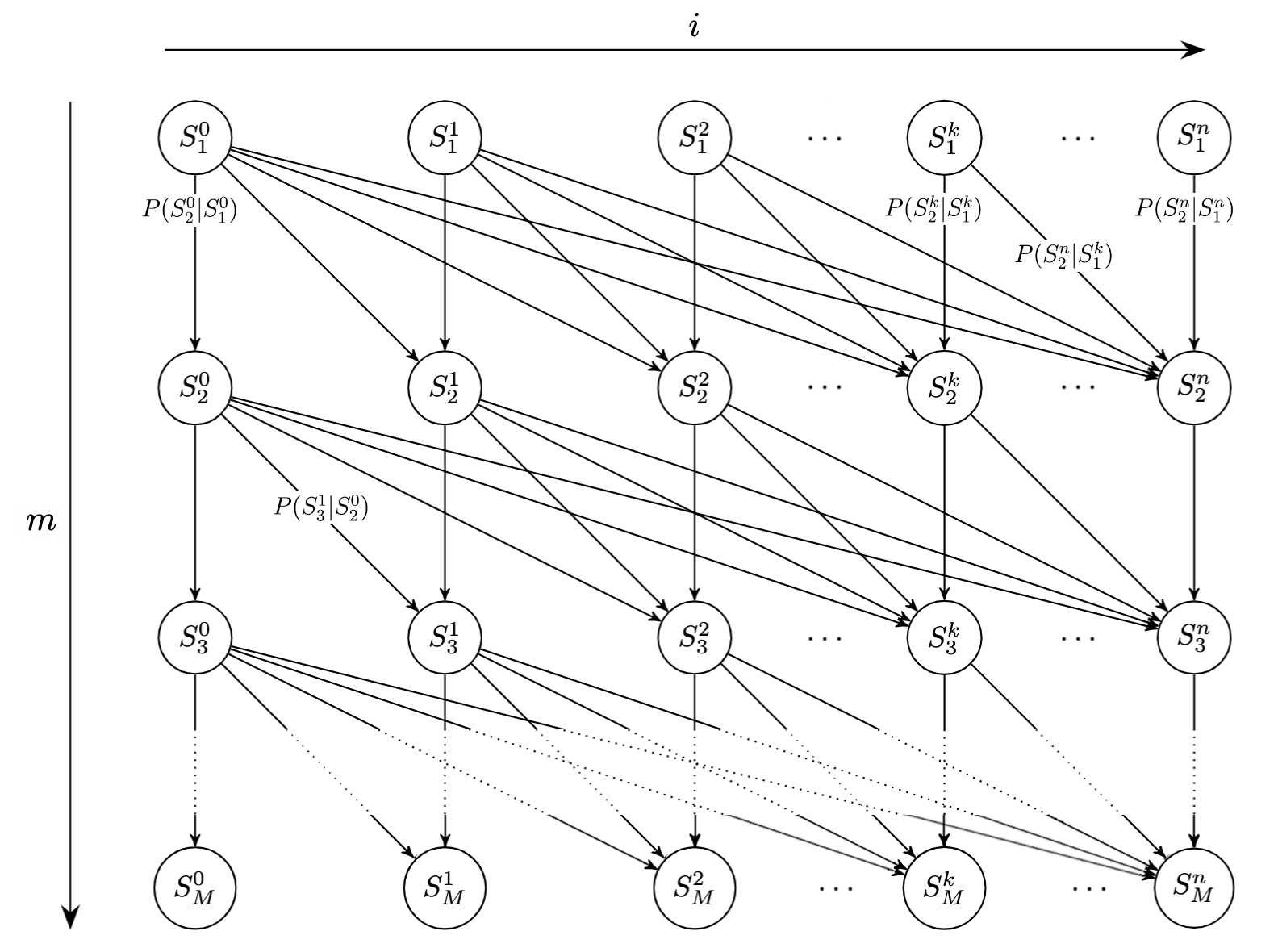}
    \caption{Markov chain for the distribution process of a multipartite entangled state. Remarkably, the problem setting only admits transitions towards states with increasing $m$ and $i$. That is, it is only possible to advance forward the Markov chain, as backward transitions are excluded by properly setting the time horizon $m$ within the coherence time.}
    \label{fig:07}
    \hrulefill
\end{figure}

It is worthwhile to note that, by adopting the proposed EAC, the identities of the contention-winning nodes are not disclosed to the other nodes. In fact, the protocol design provides each node $N_i$ with only local knowledge about the contention-resolution as a result of the measurement on the qubit of $\ket{\Lambda_{n+\ell}}$ at its side.  Differently, full knowledge about the contention-resolution is available at the orchestrator by simply measuring the ancilla qubits. 
Furthermore, in this example, we embraced the EAC with an additional characteristic. In fact, by exploiting the contention-resolution qubit measurement at each node, an EPR shared between the contention winning nodes is deterministically and distributively generated, by resorting to local operations only at the nodes. As a matter of fact, the specific generated Bell state can be determined by exploiting the properties of states $\ket{\psi}$ and $\ket{\psi'}$. Specifically, the orchestrator, that is aware of the contention-winning node identities by design, is able to resolve any ambiguity between the two states by performing a parity check on the measurement outcomes $\{g_h\}_{h \neq i,j}$ received by the nodes\footnote{Any knowledge, that may be guessed about the node identities by observing the transmissions of $\{ g_h\}_{h \neq i,j}$, can be easily obfuscated through dummy random values sent by the contention-winning nodes.}. This last observation constitutes an example of the exploitation of the fully knowledge available at the orchestrator for a certain network functionality, i.e., within the considered example for resolving the ambiguity in the extracted EPR state. 

\section{Quantum-Genuine EAC Analysis}
\label{sec:5}

In this section, we analyze the effects of noisy entanglement distribution on the proposed EAC.  To this aim, in Sec.~\ref{sec:5.1} we preliminary asses that, in presence of noise, the entanglement distribution process can be conveniently modeled with a Markov chain, and we analytically derive the closed-form expressions of both state and transition probabilities. This is a key result since Markov chain provides a powerful analytical framework that can be exploited beyond the scope of this manuscript. Then in Sec.~\ref{sec:5.2} and Sec.~\ref{sec:5.3}, stemming from the aforementioned results, we analyze the noise impact on the proposed EAC.

\subsection{Markov Chain Model}
\label{sec:5.1}

As mentioned in Sec.~\ref{sec:2.1}, the orchestrator locally generates the Dicke state $\ket{D^k_n}$ to be encoded into the contention-resolution state $\ket{\Lambda_{n+\ell}}$ reported in \eqref{eq:07}. The orchestrator than proceeds with the generation and distribution of the $n$ EPR pairs required for teleporting the qubits of the Dicke state to the $n$ contending nodes. The distribution process lasts at most $M$ time slots, where $M$ has been defined in Def.~\ref{def:07}, to account for the decoherence effect.
Specifically, in the first time slot denoted with $m=1$, $n$ heralded distribution attempts of $n$ EPR pairs are performed, one for each contending node $N_i$. By accounting for an heralded strategy, the orchestrator is able to recognize which node eventually experienced an ebit distribution failure on the quantum channel. Then, further attempts are performed over the links having experienced failure (if any) until the $M$-th time slot, which determines the time horizon of the considered system model.

By accounting for the above, we model the entanglement distribution in the distribution-attempt $m \in \{1, \ldots, M\}$ through the ``system state'' $S_m^j$, which represent the random variable associated with the connected set $\mathcal{S}_m^j$ defined in Def.~\ref{def:08}. Accordingly, the r.v. $S_m^0$ is associated with an empty connected set at time slot $m$.

\begin{prop}[Markov Condition]
    \label{prop:1}
    The sequence of random variables $S_1^h, S_2^\ell, \ldots, S_{m-1}^i, S_m^j$ is a discrete-time Markov chain, namely, it results:
        \begin{equation}
        \label{eq:15}
            P(S_m^j\mid S_{m-1}^i,\ldots,S_1^h)=P(S_m^j\mid S_{m-1}^i),
        \end{equation}
    with $i, j \in \{0,1,\ldots,N\}$ such that $j\geq i$ and $m \in \{2, \ldots, M\}$.
    \begin{IEEEproof}
        Please refer to Appendix ~\ref{app:1}. 
    \end{IEEEproof}
\end{prop}
A representation of the discrete-time Markov chain modeling the noisy entanglement distribution process is given in Figure~\ref{fig:07}, where system states are represented as nodes and allowed transitions are represented with arrows. Indeed, to evaluate the transition probabilities $\{P(S_m^j\mid S_{m-1}^i)\}_{\{m,j,i\}}$ and the state probabilities $\{P(S_m^j)\}_{\{m,j\}}$, it is useful to introduce the discrete-time stochastic vector $\textbf{X}(m)$, defined as:
\begin{align}
    \label{eq:16}
     &\textbf{X}(m)=[X_1(m),\cdots,X_n(m)],
    \,\,\, \text{with } m= 1,\cdots, M, 
\end{align}
where the component subscripts $i=1, \ldots, n$ denote the competing nodes identities. Specifically, $X_i(m) \in \textbf{X}(m)$ denotes the overall indicator variable for the $m$-distribution attempts between the orchestrator and node $N_i$.  Accordingly, $X_i(m)=0$ corresponds to the event \textit{"all the $m$-distribution attempts for node $N_i$ failed"}, while $X_i(m)=1$ corresponds to the complementary event \textit{"at least one of the $m$-distribution attempts succeeded"}. Clearly, a realization $\textbf{x}(m)=[x_1(m),\cdots,x_n(m)]$ of $\textbf{X}(m)$ has values in $\{0,1\}^{\otimes n}$. By accounting for the above and for the hypotheses in Sec.~\ref{sec:2}, the following result is obtained.
\begin{prop}
    \label{prop:2}
    The indicator random variable $X_i(m)$ for the $m$-distribution attempts between the orchestrator and node $N_i$ is characterized by the probability mass function:
    \begin{equation}
        \label{eq:17}
        P(X_i(m)=x_i(m))=
            \begin{cases}
                1-q^m, & \text{if}\, x_i(m)=1\\
                q^m,   & \text{if}\, x_i(m)=0\\
            \end{cases}
    \end{equation}
    \begin{IEEEproof}
        Please refer to Appendix~\ref{app:2}.
    \end{IEEEproof}
\end{prop}

By accounting for the result in Prop.~\ref{prop:2}, it is easy to verify that the transition probabilities $P(S_m^j\mid S_{m-1}^i)$ are given by:

\begin{align}
    \label{eq:18}
   P(S_m^j\mid S_{m-1}^i)& =\nonumber\\ P(S_m^j \mid ||\textbf{X}(m-1)||^2=i)&= \binom{n-i}{j-i} (1-q)^{j-i} q^{n-j}. 
\end{align}
Indeed, given state $S_{m-1}^i$, the probability of evolving into state $S_m^j$ -- namely, the probability of succeeding in distributing $(j-i)$ ebits -- is given by the probability of observing $(j-i)$ successes of a binomial random variable with parameters $n-i$ and $p=1-q$. From \eqref{eq:18}, it clearly results that the transition probabilities are $m$-independent.  
Furthermore, $P(S_m^j\mid S_{m-1}^i)$ is determined by the number $i$ of successful attempts achieved up to the previous slot attempt $m-1$. With the above in mind, we can now derive the following result, which characterizes the state probabilities $\{P(S_m^j)\}_{\{m,j\}}$.
\begin{cor}
    \label{cor:1}
    The probability $P(S_m^j)$ of the system being in the state $S_m^j$ at the $m$-th slot attempt follows the Binomial distribution with parameters $n$ and $1-q^m$: 
    \begin{align}
        \label{eq:19}
        P(S_m^j) &= \binom{n}{j}(1-q^{m})^jq^{m(n-j)} 
    \end{align}
    \begin{IEEEproof}
        Please refer to Appendix~\ref{app:3}
    \end{IEEEproof}
\end{cor}

\begin{figure*}[t]
    \centering
    \includegraphics[width = 0.8\textwidth]{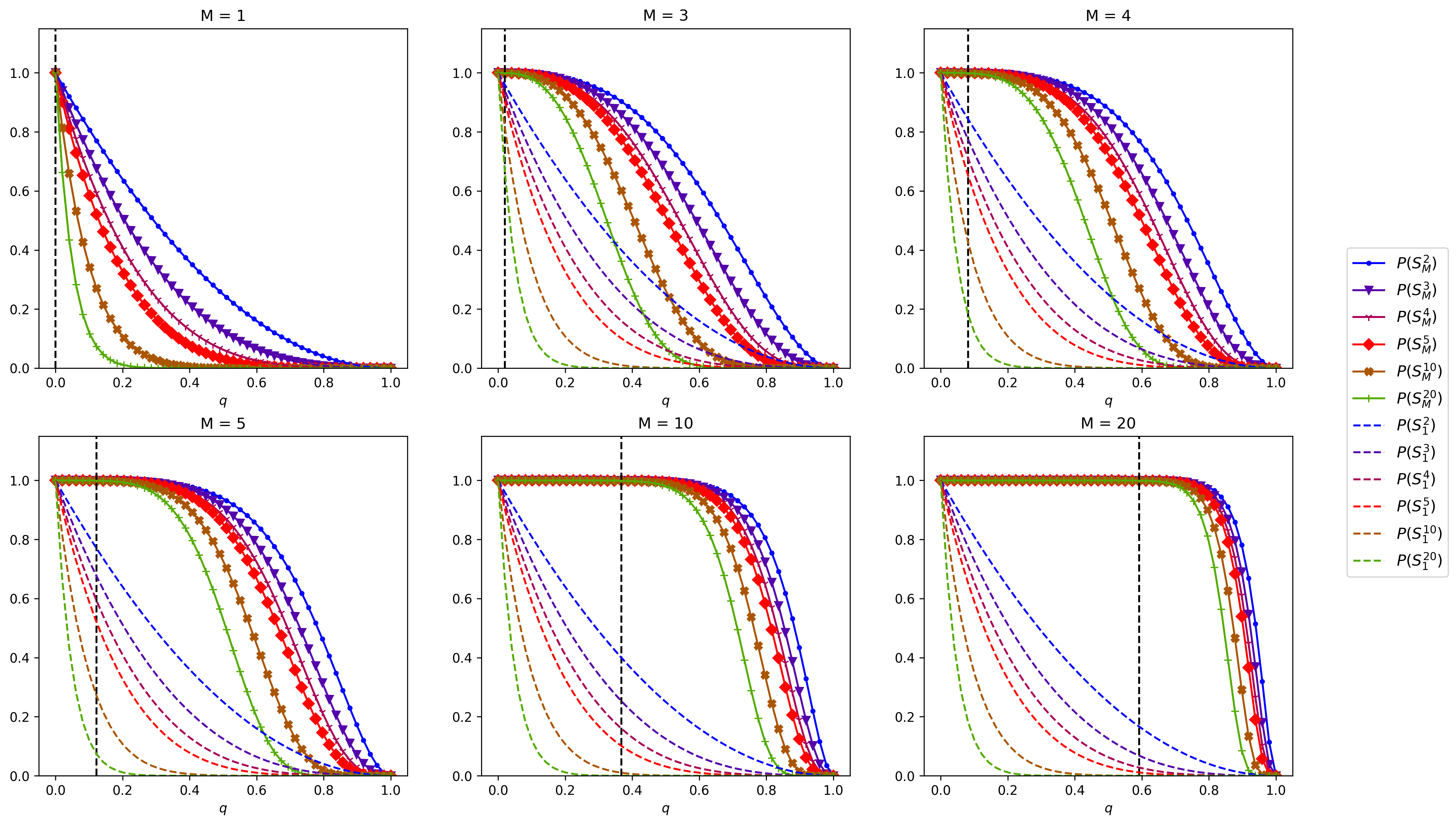}
    \caption{State probability $P(S^j_M)$, with $j=n$, versus failure distribution probability $q$, for different values of $n$ and distribution attempts $M$. The vertical dotted line represents the absorbing threshold given in \eqref{eq:23}, corresponding to a contention set of $n=20$ nodes and $\varepsilon=10^{-5}$.}
    \label{fig:08}
    \hrulefill
\end{figure*}

\subsection{Noisy Contention-Resolution State Distribution}
\label{sec:5.2}

Stemming from the analytical framework developed in the previous subsection, now we evaluate the impact of the noisy entanglement distribution on the proposed EAC. Specifically, here we conduct a theoretical analysis to understand the noise effects on the EAC when the distribution of the \textit{contention-resolution} state is noisy. Then, in Sec.~\ref{sec:5.3}, we generalize this analysis by introducing the noise also in the distribution of the \textit{contended resource}, i.e. the entanglement resource to be exploited after the contention resolution. To this aim, let $\textbf{D}=[D_1,\cdots,D_n]$ denote the random vector modeling the Dicke state $\ket{D^k_n}$ measurement outcomes, and let $\textbf{d}_h \in \{0,1\}^{\otimes n}$, with $d_{H}(\textbf{d}_h)=k$ and $h \in \{1, \ldots, \binom{n}{k}\}$, denote an arbitrary realization of $\textbf{D}$. The generic element $d_i$ of $\textbf{d}_h$ represents the outcome corresponding to the measurement of the $i$-th qubit of $\ket{D^k_n}$, which can be clearly either 0 or 1. Stemming from \eqref{eq:06}, it results:

\begin{align}
    \label{eq:20}
    P(\textbf{D}=\textbf{d}_h)&=P(\textbf{D}=\textbf{d})=\frac{1}{\binom{n}{k}} \nonumber\\\text{ with }&P(D_i=d_i)=\begin{cases}
        \frac{k}{n}&d_i=1\\
        1-\frac{k}{n}&d_i=0\\
    \end{cases}
    \; \forall \, i=1,\cdots,n 
\end{align}

Let $\mathcal{K}_{\textbf{d}_h} = \{i : d_i=1\}$ denote the identities of the $k$ nodes granted with the access to the contended resource. Each realization $\textbf{d}_h$ corresponds to a different set of $k$ selected nodes.

\begin{prop}
    \label{prop:3}
    In presence of noisy \textit{contention-resolution} distribution, the proposed EAC successfully solves the contention with probability $P_s$ given by:
    \begin{align}
    \label{eq:21}
    &P_s= (1-q^{M})^k.
\end{align}
    \begin{IEEEproof}
        Please refer to Appendix~\ref{app:4}.    
    \end{IEEEproof}
\end{prop}

\begin{rem}
    Proposition~\ref{prop:3} provides a compact closed-form expression capturing the complex behavior of the proposed EAC in presence of noise. Indeed, it results that the capability of the quantum-genuine EAC to successfully solve the contention does not depend on the total number of nodes $n$. Conversely, it depends only on the number $k$ of nodes to grant with the entanglement access as well as on the number $M$ of possible e-bit distribution attempts. This is remarkable, since random MAC protocols typically exhibit a success probability which decreases with the number $n$ of nodes sharing the medium.
\end{rem}

\begin{rem}
    An additional observation comes from the result in Proposition~\ref{prop:3}. Specifically, as observed in the previous remark, $P_s$ increases with $M$: indeed, it converges to $1$ exponentially faster as $M$ increases. This allows a further insight. Specifically, as highlighted in Section~\ref{sec:2}, we considered the worst case scenario with a single communication qubit for each resource (contenting or contention-resolution) at each node. Hence, the ebit distribution attempts are sequentially performed, and at most $M$ distribution attempts can be performed within the coherence time. Conversely, if we relax this worst-case assumption by increasing the number of communication qubits available at each network node, we can have multiple ebit distribution attempts performed in parallel, and this would directly map into higher $P_s$ as follows
        $P_s= (1-q^{M \cdot l_c})^k$,
    with $l_c$ denoting the communication qubits at each node.
\end{rem}
\begin{rem}
    Furthermore, by accounting for Proposition~\ref{prop:3} and the previous remarks, it is possible to extend the proposed EAC and its analysis to the case of the entanglement distribution enriched with \textit{entanglement purification}. \textcolor{red}{Broadly}, entanglement purification strategies consist in obtaining a single entangled state characterized by an higher fidelity from multiple imperfect entangled states\cite{BenBraPop-96}. In the context of the EAC, in order to perform entanglement purification\footnote{Hereafter, we focus on the purification of the EPR pairs needed for distributing the multipartite entangled states via teleportation. Clearly, a purification process (at the orchestrator, before the distribution process) could be required if the generation of the multipartite entangled state at orchestrator is noisy. In this case, it should be noted that an entanglement purification protocol for multipartite systems would be required \cite{YanZhoZho-23}. Such a protocol should be selected accordingly to the entangled state to be purified and to the adopted  technological implementation. However, the conducted analysis in terms of EAC success probability and distribution process does not change.}, each end-node would require more than one EPR pair for each multipartite entangled state.
     Specifically, let $\Bar{n}$ be the number of ebits required at each node to perform entanglement purification and distill one EPR with higher fidelity, set accordingly to the considered application. Then, this scenario can be analyzed by considering an equivalent system composed by $n\Bar{n}$ end-nodes without entanglement purification. Indeed, the probability $P_s$ does not depend on the cardinality of the contention set. Differently, it depends on $k$, namely, the cardinality of the subset of nodes granted with the access to the resource, which is predetermined and does not change with the adopted purification strategy. As a consequence, in case of multiple distribution attempts, the analysis developed above does not change.
    However, when parallel distribution attempts are not allowed, the purification strategy causes delays in the distribution process as it demands for additional time-slots.
\end{rem}

\begin{figure}[t]
	\centering
	\centering
	\includegraphics[width=1\columnwidth]{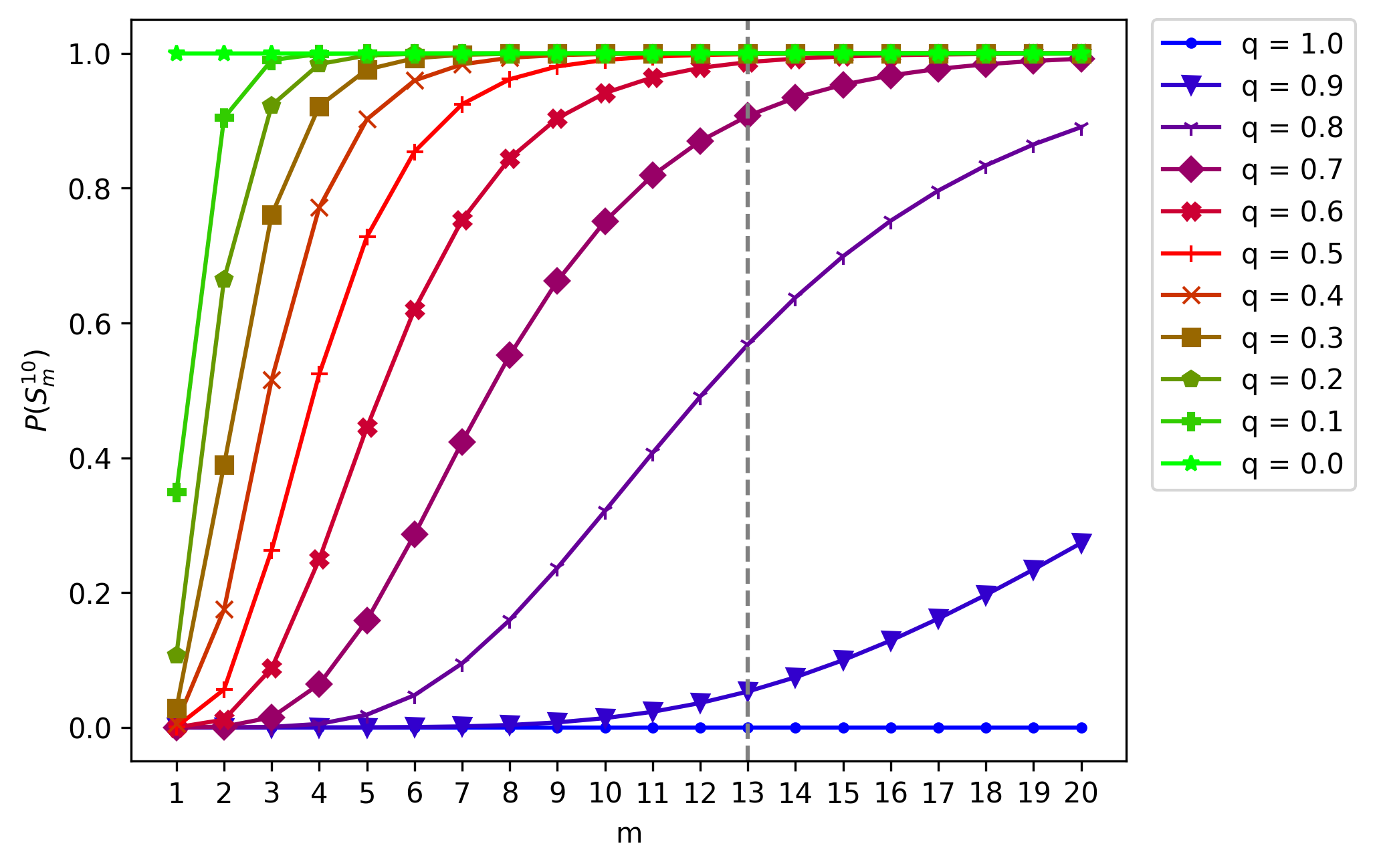}
	\caption{State probability $P(S^j_m)$, with $j=n=10$, versus the number $m$ of distribution attempts for different values of $q$.}
	\label{fig:08l}
    \hrulefill
\end{figure}

\begin{figure*}[t]
    \centering
    \includegraphics[width=0.9\linewidth]{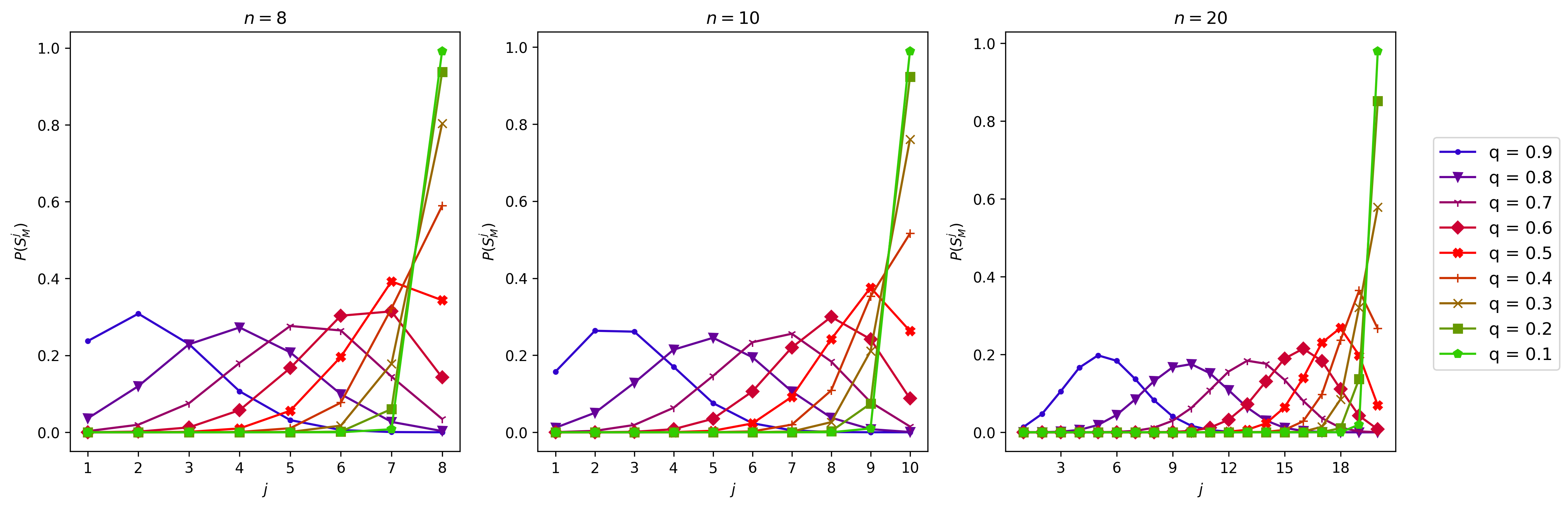}
    \caption{State probability $P(S_3^j)$ versus number $j$ of connected nodes, for different values of failure probability $q$ and node number $n$.}
    \label{fig:10}
    \hrulefill
\end{figure*}

\subsection{Fully Noisy Distribution}
\label{sec:5.3}
In the previous subsection, we considered the performance of the proposed EAC when the distribution of the \textit{contention-resolution} state is noisy. Such an analysis is here generalized by considering as ``noisy'' also the distribution of the \textit{contended resource}.  
To this aim, we observe that in the context of the EAC, the distribution of both the entangled resources could take place sequentially or in batch. In other words, the two resources may be distributed according to a time division strategy or frequency division strategy \cite{GruSaxBoa-21,NafDjo-23-1}. In both the cases, the analytical framework developed in the previous sections continue to holds. In other words, each noisy distribution process can be modeled through two independent Markov chains, driven by different failure probabilities and different time horizons $M$ in Def.~\ref{def:07}. Accordingly, in the following, we denote with $q_{\text{cr}}$ the probability of observing a failure in the distribution of the \textit{contention-resolution} state and with $q_{\text{e}}$ the probability of observing a failure in the distribution of the \textit{contended resource}. Thus, the following result is obtained. 

\begin{figure}
	\centering
	\includegraphics[width=1\columnwidth]{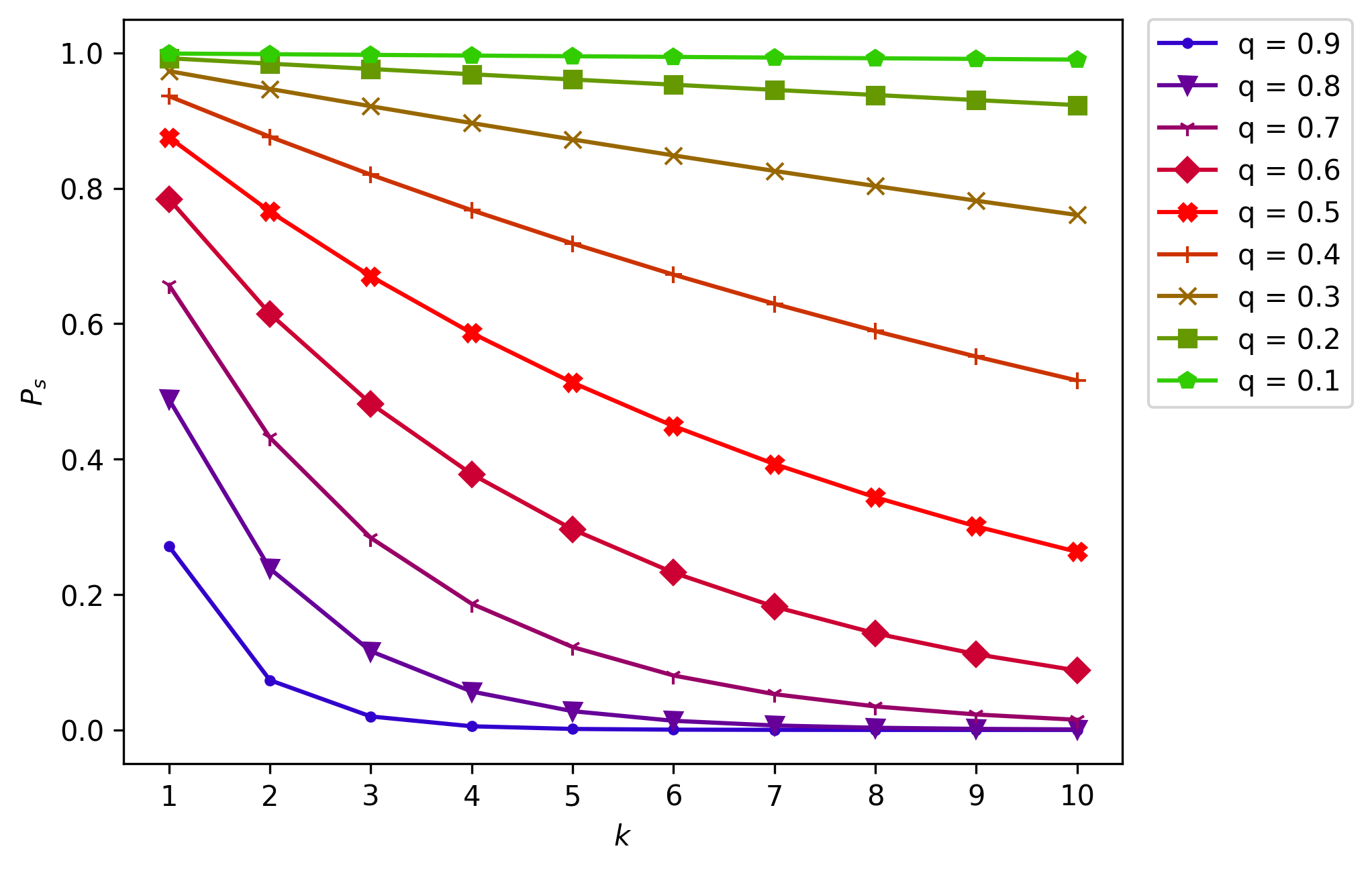}
	\caption{EAC contention-resolution probability $P_s$ versus number of nodes $k \leq n = 10$, for different values of $q$ when $M=3$.}
	\label{fig:09}
    \hrulefill
\end{figure}

\begin{prop}
    \label{prop:4}
    In presence of noisy distribution for both the resources, the proposed EAC successfully solves the contention with probability $P_s$ given by: 
    \begin{equation}
        \label{eq:23}
        P_{s} =
            (1-q_{\text{cr}}^{\overline{M}}-
            q_{\text{e}}^{\overline{M}}+
            q_{\text{cr}}^{\overline{M}}
            q_{\text{e}}^{\overline{M}})^k,
    \end{equation}
    where ${\overline{M}}$ represents the minimum time horizon characterizing the two distribution processes. 
    \begin{IEEEproof}
        Please refer to Appendix~\ref{app:5}
    \end{IEEEproof}
\end{prop}

\section{Performance Analysis}
\label{sec:6}

We first focus on the analysis of the Markovian model describing the noisy entanglement distribution. Then, we conduct a performance analysis of the EAC in presence of noise.

\subsection{Entanglement Distribution Analysis}
\label{sec:6.1}
Considering the distribution process of the \textit{contention-resolution} resource.
The probability of successfully distribute $n$ ebits, according to the Markov chain model depicted in Fig.~\ref{fig:07}, is showed in Fig.~\ref{fig:08} as a function of the failure distribution probability $q \eqdef 1 - p$ -- modeling the behavior of the physical quantum channel utilized for the entanglement distribution-- and for different values of $n$. Specifically, each plot in Fig.~\ref{fig:08} depicts: i) with a continuous line, the probability $P(S^n_M)$ of observing $n$ connected nodes after the $M$-th distribution attempt, ii) with a dotted line, the probability $P(S^n_1) = p^n$ of observing $n$ connected nodes when a single distribution attempt is performed. By considering these two metrics, we are able to compare the $M$-attempts distribution strategy with the one-shot ($M=1$) distribution strategy. As expected, in presence of noise, the distribution strategy performed over $M$ attempts outperforms the one-shot strategy. This gain is assured also in presence of hostile quantum communication channels, i.e., for the highest values of $q$. More into details, it is possible to reach $P(S^n_M)=1$ even for $q>0$. And, indeed, as shown in Fig.~\ref{fig:08}, the higher is the value of $M$, the wider is the range of $q$ values for which $P(S^n_M)=1$. Thus, the wider is the $q$ range tolerated by the distribution process. This behavior continue to hold when the number of nodes $n$ to be connected increases, although for small values of $n$ the desired $P(S^n_M)=1$ is reached in wider ranges of $q$. Stemming from this, it is evident that multiple-attempts distribution strategies should be adopted in complex quantum systems, characterized by large number $n$ of nodes and hostile quantum communication channels.
\begin{figure*}
    \centering
    \begin{subfigure}[b]{\textwidth}
    \centering
        \includegraphics[width=0.87\textwidth]{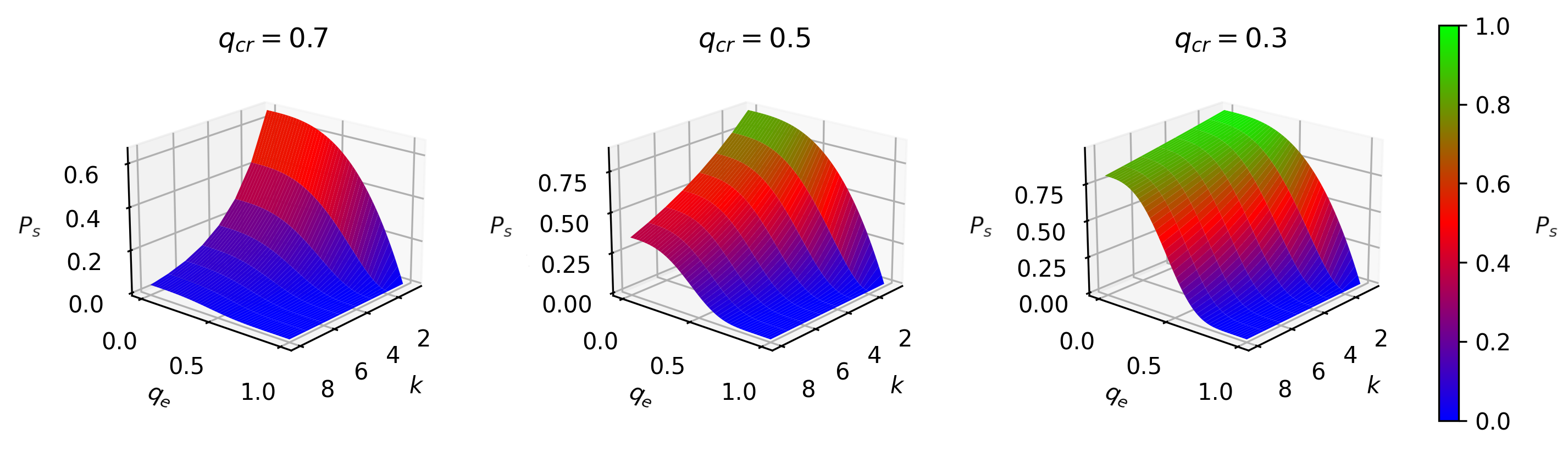}
    \end{subfigure}
    
    \begin{subfigure}[b]{\textwidth}
    \centering
        \includegraphics[width=0.87\textwidth]{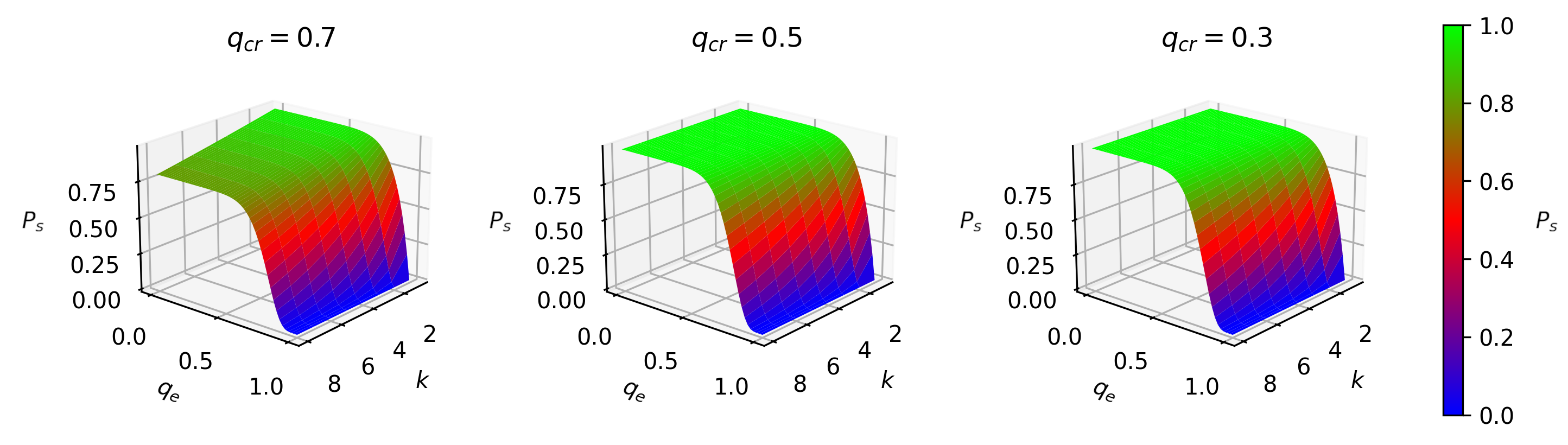}
    \end{subfigure}
        
    \caption{EAC contention-resolution probability for noisy entanglement distribution towards $n=8$ contending nodes, with $\overline{M}=3$ for the uppermost figures and $\overline{M}=10$ for the lowermost figures.}
    \label{fig:11x}
    \hrulefill
\end{figure*}
To elaborate more on the above considerations, we can observe that for any value of $M$, each subplot in Fig.~\ref{fig:08} shows a range of $q$-values corresponding to a region of $P(S_M^n)\approx1$ and a complementary range of $q$ characterized by appreciable variations of $P(S_M^n)$. As instance for $M=20$, the probability of having $n$ connected nodes changes with $n$ only in presence of hostile quantum communication channels, i.e., for $q \in (0.6,1]$, although the curves corresponding to different $n$ exhibit higher slopes and they are close to each others. Conversely, for $q \leq 0.6$, the curves corresponding to different numbers $n$ of connected nodes are indistinguishable. Thus, higher values $M$ of attempts reduces the $q$-range significantly impacting on the performances of the entanglement distribution processes and, consequently, on the proposed EAC. Differently, as the value of $M$ decreases, the probability $P(S_M^n)$ of observing $n$ connected nodes after $M$-attempts exhibits a heterogeneous behaviour as the number of nodes $n$ to be connected varies, with a ``shrinking'' of the $q$-range in which $P(S_M^n) \approx 1$.
Stemming from this, the number $M$ of attempts determines the so called \textit{absorbing threshold} $\Bar{q}(M)$, represented with a dotted vertical line in Fig.~\ref{fig:08}, i.e., the value of the probability $q$ such that:
\begin{equation}
    \label{eq:24}
    P(S_M^n) > 1 - \varepsilon, \quad  \forall q < \Bar{q}(M)
\end{equation}
for any considered\footnote{The absorbing threshold depends on $n$ as well, i.e., for $n_1>n_2: \Bar{q}_{n_1}(M)<\Bar{q}_{n_2}(M)$. Hence, to provide a global parameter meaningful for all the considered values of $n$, we considered in Fig.~\ref{fig:08} the worst-case scenario, with the threshold corresponding to the highest value of $n$. The dependence on $n$ appears to have weaker effects on the threshold than the one caused by $M$.} value of $n$. In Fig.~\ref{fig:08}, $\epsilon$ is set equal to $10^{-5}$. 
The above considerations can be further extended by considering the probability $P(S^j_m)$ of observing $j=n$ connected nodes as a function of the number $m$ of distribution attempts, as depicted in Fig.~\ref{fig:08l}, for $n=10$. As showed by the line for $q=0.4$, $P(S^n_m)\approx 1$ after $m=13$ attempts. Clearly, a lower number of attempts is required to reach the value $P(S^n_m)\approx 1$ for $q<0.4$. Finally, in Fig.~\ref{fig:10}, we report the state probability $P(S_M^j)$ as a function of the number $j$ of connected nodes, for different values of node number $n$ and different values of failure probability $q$. $P(S_M^j)$ represents the probability of observing $j$ out of $n$ connected nodes at the $M$-th attempt, namely, the probability of observing the system in one of the states belonging to the last row of the Markov chain depicted in Fig.~\ref{fig:07}. According to Corollary~\ref{cor:1}, for a given number of attempts ($M=3$ in Fig.~\ref{fig:10}) the connected set probability follows a binomial distribution with parameters $n,1-q^M$. It is worthwhile to note that the grainy curves are due to undersampling, as a consequence of the small values of $j$. From Fig.~\ref{fig:10}, for a given $q$, it results that when the number $n$ of nodes increases $P(S_M^j)$ decreases.  

\subsection{EAC Performance Analysis}
\label{sec:6.2}

We focus now on investigating the performance of the proposed quantum-genuine EAC in presence of noise.
To this aim, in Fig.~\ref{fig:09}, we report the contention-resolution probability $P_s$ derived in Prop.~\ref{prop:3} as a function of the number $k \leq n$ of nodes to be granted with access to the contending resource, for different values of failure distribution probability $q$. The number $M$ of distribution attempts is set equal to $M=3$. As showed within the figure, and in agreement with equation~\eqref{eq:21}, the contention-resolution probability decreases with the number $k$ of nodes to be granted with access to the contended resource, for a given value of $q$.
The above analysis can be generalized by accounting for the noise also in the distribution of the \textit{contended} entangled resource as derived in Prop.~\ref{prop:4}. Specifically, in Fig.~\ref{fig:11x}, the contention-resolution probability \eqref{eq:23} is reported for different communication settings by considering different ebit loss probabilities for the contended resource and the contention-resolution one. 

Accordingly to the analysis developed at the beginning of this section, in order to emphasize the noise effects on the proposed EAC, we consider a low number $\overline{M}$ of attempts. More into details, in the uppermost figures of Fig.~\ref{fig:11x}, $P_{s}$ is numerically evaluated for $\overline{M}=3$, when the contention set is constituted by $n=8$ nodes. 
As expected, the overall performances are significantly lowered by the noisy distribution of both the entangled resources. As instance, with reference to Fig.~\ref{fig:09}, the contention-resolution probability for $q_{\text{cr}}=0.3, q_{\text{e}}=0$ and $k=2$ reaches values close to one. Differently, Fig.~\ref{fig:11x} shows significantly lower values of the contention-resolution probability. And indeed this $P_s$ decrease is more significant for higher values of $q_{\text{e}}$. The reason is twofold. On one hand, high values of $q_{\text{e}}$, that is related to the channel absorption experienced by the distribution of the contended resource, prevents to reach an high number of nodes. On the other hand, the identities of the nodes granted with the access to the resource also impact the contention-resolution probability. Indeed, as showed in the uppermost figures of Fig.~\ref{fig:11x}, even when both the resources experience low absorption, say $q_e=q_{cr}=0.3$, $P_s$ has maximum value equal to $0.8$ for $k=2$. 
As showed in Fig.~\ref{fig:11x}, higher values of $P_s$ can be achieved with an higher number of distribution attempt, i.e., higher values of $\overline{M}$. Specifically, the lowermost figures show the contention-resolution probability for a system supporting $\overline{M}=10$ attempts for the distribution of the entangled resources, towards $n=8$ contending nodes. We observe that the contention-resolution probability exhibits quite the same behavior in presence of different channel conditions, namely $q_{cr}=0.5$ and $q_{cr}=0.3$. Thus, multiple-distribution attempts can be an effective strategy to overcome different noise severity levels. Indeed, for $q_e=q_{cr}=0.7$, the contention-resolution probability shifts from a maximum value about $0.1$ when $\overline{M}=3$ to a maximum value in $[0.75,1)$ when $\overline{M}=10$.

\section{Conclusions}
\label{sec:7}
We proposed a quantum-genuine EAC protocol which leverages the features of the multipartite entangled Dicke states to accomplish the access control of a contended entangled resource. Such a protocol preserves the privacy and anonymity of the identities of the selected nodes, and it avoids to delegate the signaling arising with entanglement access control to the classical network. Additionally, we conducted a theoretical analysis of the EAC in presence of noise. Specifically, we developed an analytical framework leveraging the features of Markov chains to capture the behavior of the entanglement distribution process in presence of noise. Stemming from this, we derived closed-form expressions of the contention-resolution probability in presence of noise. This theoretical analysis is able to catch the complex noise effects on the EAC through meaningful parameters. This, in turn, provides crucial guidelines for the design of efficient network protocols. Finally, a numerical analysis is conducted, which shows that the noise effects are narrowed through multiple-attempt distribution strategies.

\begin{appendices}

\section{Proof of Proposition~\ref{prop:1}}
\label{app:1}

The proof easily follows by observing that the probability of evolving to state $S_m^j$, starting from the state evolution $S_1^h, \ldots, S_{m-1}^i$, depends only on the current state $S_{m-1}^i$. In fact, by accounting for the system model given in Sec.~\ref{sec:2}, the heralded teleportation scheme allows the orchestrator to recognize which node -- if any -- experienced an ebit loss. In such a case, further distribution attempts can be performed to eventually distribute the targeted state to the missing nodes. Furthermore, by restricting the distribution attempts within the time horizon $M$, namely, within a time interval where the decoherence effects are negligible, the system state evolution is restricted from ``backward'' transitions towards smaller connected sets with $j<i$ . As a consequence, once in the state $S_{m-1}^i$, the orchestrator performs $n-i$ new generation and distribution attempts trying to reach the missing contending nodes.

\section{Proof of Proposition~\ref{prop:2}}
\label{app:2}
    
According to Def.~\ref{def:06}, each distribution attempt is modeled as a Bernoulli random variable with parameter $p$. Accordingly, $q\eqdef1-p$ denotes the probability of observing a failed distribution attempt. Once node $N_i$ experienced a distribution success, no further distribution attempts are performed. Hence, $X_i(m)=0$ \textit{iff} exactly $m$ distribution attempts failed. This event occurs with probability $q^m$. And, accordingly, the corresponding complementary event has probability $1-q^m$. Equivalently, this can be expressed by stating that $X_i(m)$ has a Bernoulli distribution with parameter $1-q^m$, i.e., $X_i(m)\sim\mathcal{B}(1,1-q^m)$. The aforementioned probability mass function holds regardless of the node identity $i$. As a consequence, $X_1(m),\cdots,X_n(m)$ are $n$ \textit{i.i.d.} Bernoulli random variables with parameter $1-q^m$.

\section{Proof of Corollary~\ref{cor:1}}
\label{app:3}

The proof follows by accountiong for the results derived in Prop.~\ref{prop:2}.
 Specifically, we observe that the indicator random variable $X_i(m)$ takes value in $\{0,1\}$. Hence, $\parallel \textbf{X}(m)\parallel^2=\sum_{i=1}^nX_i(m)$. As a consequence, it results 
\begin{align}
\label{eq:a.01}
    P(S_m^j)&=P(\parallel \textbf{X}(m)\parallel^2=j) =P(\sum_{i=1}^nX_i(m)=j).
\end{align}
Being $X_1(m),\cdots,X_n(m)$ \textit{i.i.d.} Bernoulli random variables with parameter $1-q^m$, as proved in Prop~\ref{prop:2}, and by exploiting the last equality in \eqref{eq:a.01}, the result follows. In fact,  the sum of $n$ i.i.d. Bernoulli random variables with success probability $1-q^m$ is distributed according to a binomial distribution with parameters n and $1-q^m$.

\section{Proof of Proposition \ref{prop:3}}
\label{app:4}

Having no less than $k$ connected nodes, i.e., $\parallel\textbf{X}(M)\parallel^2\geq k$, is a necessary condition for the success of the contention resolution. However, such a condition is not a sufficient condition. Specifically, given the set of identities $\mathcal{K}_{\textbf{d}_h}$ corresponding to a certain realization $\textbf{d}_h$, it must also hold that an e-bit has been successfully distributed to the nodes with identities in $\mathcal{K}_{\textbf{d}_h}$. In other words, the winning set must belong to the connected set $\mathcal{S}^j_M$, i.e., $\mathcal{K}_{\textbf{d}_h} \subseteq \mathcal{S}^j_M$, or equivalently $X_i(M) = 1 , \forall i \in \mathcal{K}_{\textbf{d}_h}$. Hence, it follows that the probability of successfully solving the contention can be expressed as:

\begin{align}
    \label{eq:a.03}
    &P_s =P(\textbf{X}(M)\textbf{D}^t = k)= \nonumber\\ &P(\textbf{X}(M)\textbf{D}^t=k \mid \, \parallel\textbf{X}(M)\parallel^2\geq k)P(\parallel\textbf{X}(M)\parallel^2\geq k) 
\end{align}
By exploiting the law of total probability, \eqref{eq:a.03} can be re-written as:

\begin{align}
    \label{eq:a.05}
    P_s  = \sum_{h=1}^{\binom{n}{k}}P(\textbf{X}(M)\textbf{d}_h^t=k\mid\parallel\textbf{X}(M)\parallel^2\geq k,\textbf{D}=\textbf{d}_h) \nonumber \\ P(\textbf{D}=\textbf{d}_h)P(\parallel\textbf{X}(M)\parallel^2 \geq k)= \nonumber \\
        =\sum_{h=1}^{\binom{n}{k}}\sum_{j=k}^{n}P(\textbf{X}(M)\textbf{d}_h^t=k\mid\parallel\textbf{X}(M)\parallel^2=j,\textbf{D}=\textbf{d}_h) \nonumber\\P(\parallel\textbf{X}(M)\parallel^2 =j) P(\textbf{D}=\textbf{d}_h)
\end{align}
By accounting for the structure of the contention-resolution state, it is easy to verify that:

\begin{align}
    \label{eq:a.06}
    P(\textbf{X}(M)\textbf{d}_h^t=k\mid\parallel\textbf{X}(M)\parallel^2=j,\textbf{D}=\textbf{d}_h)=\nonumber \\ \binom{n-k}{j-k} / \binom{n}{j}, \forall j \geq k
\end{align}
which does not depend on the particular $\textbf{D}=\textbf{d}_h$. Thus, it results that the conditioning on $\textbf{D}=\textbf{d}_h$ in \eqref{eq:a.05} does not act. Consequently, \eqref{eq:a.05} can be re-written as follows:

\begin{align}
    \label{eq:a.07}
    P_s =\sum_{j=k}^{n}P(\textbf{X}(M)\textbf{d}^t=k\mid\parallel\textbf{X}(M)\parallel^2=j,\textbf{D}=\textbf{d}) \nonumber\\ P(\parallel\textbf{X}(M)\parallel^2 =j). 
\end{align}
By accounting for the result in Corollary~\ref{cor:1} and by 
substituting \eqref{eq:a.06} in \eqref{eq:a.07}, one obtains:

\begin{align}
    \label{eq:a.08}
    P_s = \sum_{j=k}^{n}\binom{n-k}{j-k} (1-q^M)^j q^{M(n-j)}=\nonumber \\  (1-q^M)^k \sum_{h=0}^{n-k}\binom{n-k}{h} (1-q^M)^h q^{M(n-k-h)}.
\end{align}
with \eqref{eq:a.08} equal to $(1-q^M)^k$ by applying the Binomial Theorem.

\section{Proof of Proposition~\ref{prop:4}}
\label{app:5}

Let $\textbf{Y}(m)$ denote the discrete-time stochastic vector associated to the distribution process of the contended resource, i.e., $\textbf{Y}(m)=[Y_1(m),\cdots,Y_n(m)]$ with $m= 1,\cdots, M_{\text{e}}$. It is worthwhile to observe that threshold time in Def.~\ref{def:02} might depend on the technology adopted for generating the multipartite entangled state. Hence, in the general case of considering heterogeneous systems, $\tau_{\text{th}}$ could be different for the two entangled resources. This, in turn, determines different values for the time horizon in Def.~\ref{def:07}. In order to evaluate the EAC successful probability, we consider $\overline{M}\eqdef \min\{M_{\text{cr}}, M_{\text{e}}\}$, with $M_{\text{cr}}$ denoting the time horizon of the \textit{contention-resolution} state distribution, and $M_{\text{e}}$ denoting the time horizon of the \textit{contended resource} distribution. With this in mind, the proofs follows by reasoning similarly to Prop.\ref{prop:3}, by observing that the success probability can be expressed as: $P_{s} = P(\textbf{X}(\overline{M})\textbf{D}^T=k,\textbf{Y}(\overline{M})\textbf{D}^T=k)=P(\textbf{X}(\overline{M})\textbf{D}^T=k)P(\textbf{Y}(\overline{M})\textbf{D}^T=k)$,

being the distribution processes of the two entangled resources statistically independent. By accounting for the result in Prop.~\ref{prop:3}, after some algebraic manipulations, the proof follows.

\end{appendices}

\bibliographystyle{IEEEtran}
\bibliography{biblio.bib}

\end{document}